\documentclass[floatfix,%
 reprint,
 amsmath,amssymb,
 aps,
pra,
]{revtex4-2}
\usepackage{graphicx,amsthm}
\usepackage{dcolumn}
\usepackage{bm}
\usepackage{xcolor}
\usepackage{comment}

\begin{document}

\preprint{APS/123-QED}

\title{Sensitivity-Driven Migration and the Evolution of Cooperation in Multi-Player Games on Structured Populations
}

\author{Dhaker Kroumi}
\email{dhaker.kroumi@kfupm.edu.sa}
 \affiliation{Department of Mathematics, King Fahd University of Petroleum and Minerals, Dhahran, S.A.}

\date{\today}


\begin{abstract}
Cooperation often depends on individuals avoiding exploitation and interacting preferentially with other cooperators. We explore how context-dependent migration influences the evolution of cooperation in spatially structured populations. Individuals interact in small groups through public goods games and reproduce with possible dispersal. Cooperators migrate more frequently when surrounded by defectors, while defectors disperse uniformly. This behavioral asymmetry reflects realistic differences in mobility and social responsiveness. Our results show that conditional migration can promote cooperation by enabling cooperators to escape defector-rich environments and cluster together. The effectiveness of this mechanism depends on baseline migration rates, group size, and the sensitivity of cooperators to local conditions. We identify parameter ranges where cooperation is favored even under conditions that would typically hinder its evolution. These findings highlight how behavioral plasticity and dispersal strategies can interact with population structure to support the emergence of cooperation.
\end{abstract}

\maketitle




\section{Introduction}
The evolution of cooperation remains a central challenge in evolutionary biology. Cooperative behavior, where individuals incur a cost to benefit others, is vulnerable to exploitation by defectors. Yet, cooperation is widespread across taxa, from microbes to humans. This apparent paradox has led to the development of several theoretical frameworks including kin selection \cite{H1964}, reciprocal altruism \cite{T1971}, indirect reciprocity \cite{NS1998}, group selection \cite{W1975}, and network reciprocity \cite{OHLN2006}, which emphasize the role of repeated interactions, relatedness, and structured populations.

While two-player games like the Prisoner’s Dilemma provide foundational insights \cite{AH1981}, many social interactions involve larger groups. Multi-player games such as Public Goods Games (PGGs), where cooperators contribute to a shared resource,  better capture such dynamics. These models introduce features like synergy \cite{HDHS2002}, threshold effects \cite{PPS2009}, and diminishing returns \cite{A2009}, which lead to complex outcomes including bistability and coexistence \cite{HD2004}. Maintaining cooperation in such settings often requires stronger assortment or conditional strategies \cite{SSP2008, SF2007, PS2010}.

Population structure facilitates cooperation by allowing cooperators to cluster and avoid exploitation. Theoretical models show that limited dispersal \cite{VR1998}, neighborhood-based interactions \cite{LHN2005}, and local competition can significantly alter evolutionary dynamics. Empirical studies in microbial colonies \cite{NXF2010}, animal societies \cite{WGG2007}, and human networks \cite{SPL2006} confirm that spatial and social structure frequently supports cooperation. The island model \cite{W1931}, especially when coupled with the Moran process \cite{M1958, L1973}, provides a tractable framework for analyzing structured populations. This setting has been used to study fixation probabilities under neutrality \cite{E2004}, weak selection \cite{EN1980}, and strategic interactions \cite{AND2013, CA2016}, as well as the influence of migration and demographic structure \cite{R2004, W2005}. Migration plays a dual role: while it can disrupt cooperative clustering, it also allows for dynamic assortment when migration is context-dependent.

Several studies suggest that cooperators can migrate strategically in response to local conditions. Mechanisms such as kin-biased dispersal \cite{GW2006}, conditional emigration \cite{A2004, A2011}, and adaptive relocation \cite{ISSW2013, JCLG2017} demonstrate the evolutionary significance of behavioral responsiveness. This form of migration increases assortment without requiring genetic relatedness or repeated interactions and may solve the cooperation dilemma in one-shot, multi-player settings. Despite strong simulation-based support for conditional migration \cite{PS2002, A2011}, few analytical models explore this behavior within evolutionary game frameworks. Moawad \textit{et al.} \cite{MAB2024} showed that cooperation fails in structured populations if migration is independent of social context, emphasizing the need for models incorporating adaptive dispersal.

In this study, we introduce a model that integrates multi-player games, an island Moran framework, and context-dependent migration. Cooperators disperse at a rate that increases with defector frequency in their natal group, while defectors migrate uniformly. This asymmetry reflects biologically realistic differences in mobility and responsiveness. Using a diffusion approximation valid under weak selection and large deme numbers \cite{EN1980, E2004}, we derive an analytical expression for the fixation probability of a single cooperator. Our results show that differential migration can significantly reduce the threshold benefit-to-cost ratio needed for cooperation to evolve. When cooperators are highly sensitive to defection and baseline migration is limited, cooperation is strongly favored even in conditions where it would fail under classical assumptions. However, when sensitivity is low, the advantage of migration vanishes, echoing recent findings in tolerance-based models \cite{PABS2023}. Our model connects a static group structure with dynamic behavioral responses, and aligns with findings from cooperation on graphs \cite{MAB2024}, dynamic social networks \cite{SMP2023}, and ecological models with local competition \cite{RBR2023}.


\section{Methods \label{sec1}}


\subsection{\label{sec1-1}Model}
Consider a structured haploid population partitioned into \( d \) demes, each containing \( N \geq 2 \) individuals. Each individual can be of type \( C \) (cooperator) or type \( D \) (defector), and interactions occur within demes or groups of size \( k+1 \), drawn randomly without replacement. The payoffs are determined by the matrix
\[
\begin{array}{c|cccc}
   & 0 & 1 & \cdots & k \\ \hline
C & a_0 & a_1 & \cdots & a_k \\
D & b_0 & b_1 & \cdots & b_k \\
\end{array}
\]
where \( a_n \) (resp. \( b_n \)) denotes the payoff to a cooperator (resp. defector) interacting with \( n \) cooperators among the \( k \) other group members. Then, under uniform random sampling, the expected payoffs for a cooperator and a defector in a deme containing $i$ cooperators and $N-i$ defectors are given, respectively, by
\begin{subequations}\label{sec1-eq1}
    \begin{align}
        \omega(C,i) &= \sum_{n=0}^{k} \frac{\binom{i-1}{n} \binom{N-i}{k-n}}{\binom{N-1}{k}} a_n, \\
        \omega(D,i) &= \sum_{n=0}^{k} \frac{\binom{i}{n} \binom{N-i-1}{k-n}}{\binom{N-1}{k}} b_n.
    \end{align}
\end{subequations}
Here, $\binom{i-1}{n}\binom{N-i}{k-n}/\binom{N-1}{k}$ is the probability of encountering exactly $n$ cooperators among the $k$ partners in the interacting group.

Selection occurs via a reproduction-migration process. At each time step, an offspring is produced in a deme chosen at random by an individual selected with a probability proportional to its fitness given by \( f = 1 + s\omega \). Here \( s =\delta/(Nd) \) denotes weak selection, a regime typical in finite populations \cite{N2006}.

The offspring produced may migrate with probability 
\begin{equation}\label{sec1-eq2}
m(X,y) = m + s M(X,y) + o(s),
\end{equation}
which depends on its strategy \( X \in \{C, D\} \) and the frequency of cooperators \( y \)parent's in the deme.
This reflects a baseline migration probability $m$ and a type-dependent correction term $M(X,y)$, consistent with first-order approximations in population models \cite{R2004,W2005}.
If the offspring migrates, it selects a deme at random and replaces an individual randomly chosen within that deme. Otherwise, it replaces a randomly chosen individual within the parent’s deme, including the parent itself.

\subsection{Approximation by a diffusion process and the fixation probability}

To capture the metapopulation-level dynamics, we define the process $\{ X(t) \}_{t \in \mathbb{N}}$, which tracks the overall frequency of cooperators across generations. For an evolutionary process under selection in a finite number of demes, the dynamics can be approximated by a limiting process in an infinite population subdivided into infinitely many demes. More precisely, in the limit of a large number of demes \( d \), the rescaled process $\{ X(\lfloor N^2 d^2 \tau \rfloor) \}_{\tau\geq 0}$, which tracks the global frequency of cooperators, converges in distribution to a continuous-time diffusion process \( \{ X^*(\tau) \}_{\tau \geq 0} \) on \([0,1]\). This convergence is established via the two-timescale framework developed in \cite{EN1980} and extended to structured populations into many demes in \cite{WT2004, KL2025}.
The limiting diffusion process is described by the generator  
\begin{equation}\label{sec1-eq3}
\mathcal{L} = \delta\mu(x) \frac{d}{dx} + \frac{\sigma^2(x)}{2} \frac{d^2}{dx^2},
\end{equation}  
where the drift and diffusion terms are defined as  
\begin{subequations}\label{sec1-eq4}
\begin{align}
\mu(x) &= \delta\sum_{i=0}^N v_i(x) x_i (1 - x_i) \big( \omega(C,i) - \omega(D,i) \big) \label{sec1-eq10} \\
&\quad + \delta\sum_{i=0}^N v_i(x) (x_i - x) \big( x_i M(C,x_i) + (1 - x_i) M(D,x_i) \big), \nonumber \\
\sigma^2(x) &= 2(1 - m) \sum_{i=0}^N v_i(x) x_i (1 - x_i) + 2 m x (1 - x). \label{sec1-eq11}
\end{align}
\end{subequations}
Here \(v_i(x)\) denotes the probability that a randomly chosen deme in a population divided into infinitely many demes contains exactly $i$ cooperators when the overall cooperator frequency is $x$, where each individual has baseline fitness $1$ and migrates between demes with probability $m$. This probability is given by the Beta-Binomial distribution as
\begin{equation}\label{sec1-eq5}
v_i(x) = \binom{N}{i} \frac{B\left(\frac{Nmx}{1-m}+i,\frac{Nm(1-x)}{1-m} + N-i\right)}{B\left(\frac{Nmx}{1-m},\frac{Nm(1-x)}{1-m}\right)},
\end{equation}  
where \( B(a,b) = \int_0^1 t^{a-1}(1 - t)^{b-1} dt \) denotes the Euler Beta function. The distribution $\mathbf{v}(x)=(v_i(x))_{i=0}^{N}$ captures the quasi-equilibrium configuration of deme compositions under neutrality, conditioned on the global cooperator frequency \( x \).

Following classical diffusion theory \cite{E2004}, the fixation probability of a single cooperator $C$ introduced into a population initially composed of defectors and subdivided into many demes of size $N$ can be approximated as
\begin{equation}\label{sec1-eq6}
\rho_C \approx \frac{\int_0^{1/(Nd)} \psi(y) dy}{\int_0^1 \psi(y) dy},
\end{equation}
where
\begin{equation}\label{sec1-eq7}
\psi(y) = \exp\left(-2 \int_0^y \frac{\mu(x)}{\sigma^2(x)} dx\right). 
\end{equation}
For $\delta<<1$, a linear approximation gives
\begin{equation}\label{sec1-eq8}
\begin{split}
\rho_C &\approx \frac{1}{Nd} + \frac{2}{Nd} \int_0^1 \int_0^y \frac{\mu(x)}{\sigma^2(x)} dxdy\\
&\approx \frac{1}{Nd} + \frac{1}{Nd} \int_0^1 2(1 - x)\frac{\mu(x)}{\sigma^2(x)} dx. 
\end{split}
\end{equation}


\section{Large deme size\label{sec2}}

For simplicity, suppose that $N$ is sufficiently large and that $m\to0$ in such a way that $\nu=Nm$ remains constant. Under these conditions, the expected payoffs for a cooperator and a defector become, respectively,
\begin{subequations}\label{sec2-eq1}
    \begin{align}
        \omega(C,i)&\approx\sum_{n=0}^{k}\binom{k}{n}(i/N)^n(1-i/N)^{k-n}a_n,\\
        \omega(D,i)&\approx\sum_{n=0}^{k}\binom{k}{n}(i/N)^n(1-i/N)^{k-n}b_n.
    \end{align}
\end{subequations}
 This is a consequency of the asymptotic estimate
 $$\frac{\binom{i-1}{n} \binom{N-i}{k-n}}{\binom{N-1}{k}}\approx\binom{k}{n}(i/N)^n(1-i/N)^{k-n}+o(1).$$
In addition, note that the discrete distribution $(v_i(x))_{i=0}^{N}$ converges to a continuous Beta distribution on $(0,1)$ with density
\begin{equation}\label{sec2-eq2}
g(y)=\frac{y^{\nu x-1}(1-y)^{\nu(1-x)-1}}{B\left(\nu x,\nu(1-x)\right)}.
\end{equation}
See Appendix D in Kroumi and Lessard \cite{KL2025} for more details. Inserting the approximations in Eqs. (\ref{sec2-eq1}) and (\ref{sec2-eq2}) into Eq. (\ref{sec1-eq4}), the drift and diffusion functions evaluated at the quasi-equilibrium $\mathbf{v}(x)$ will take the form
\begin{equation}\label{sec2-eq4}
\begin{split}
&\mu(x)=\delta\sum_{i=0}^{N}v_i(x) x_i (1 - x_i) \left(\omega(C,i) - \omega(D,i)\right)\\
&+ \delta\sum_{i=0}^{N}v_i(x)  (x_i - x) \left(x_iM(C,x_i)+(1-x_i)M(D,x_i)\right), \\
&\approx\delta\sum_{n=0}^{k}\binom{k}{n}\int_{0}^{1}y^{n+1}(1-y)^{k+1-n}g(y)dy\left(a_n-b_n\right)\\
&+\delta\int_{0}^{1}(y-x)\left(yM(C,y)+(1-y)M(D,y)\right)g(y)dy\\
&\approx\delta\sum_{n=0}^{k}\binom{k}{n}\frac{B\left(\nu x+n+1,\nu(1-x)+k+1-n\right)}{B\left(\nu x,\nu(1-x)\right)}\left(a_n-b_n\right)\\
&+\delta\int_{0}^{1}(y-x)\left(yM(C,y)+(1-y)M(D,y)\right)g(y)dy
\end{split}
\end{equation}
and 
\begin{align}
\sigma^2(x)&=2(1-m)\sum_{i=0}^{N}v_i(x)x_i(1-x_i)+2mx(1-x)\nonumber\\
&\approx 2\int_{0}^{1}y(1-y)g(y)dy\nonumber\\
&\approx2\frac{B\left(\nu x+1,\nu(1-x)+1\right)}{B\left(\nu x,\nu(1-x)\right)}\nonumber\\
&\approx\frac{2\nu}{\nu+1}x(1-x).\label{sec2-eq5}
\end{align}
Here, we have used the facts that $B(x+1,y)=\frac{x}{x+y}B(x,y)$ and $B(x,y)=B(y,x)$.

\begin{figure}
\centering
\includegraphics[height=6.5cm, width=8.7cm]{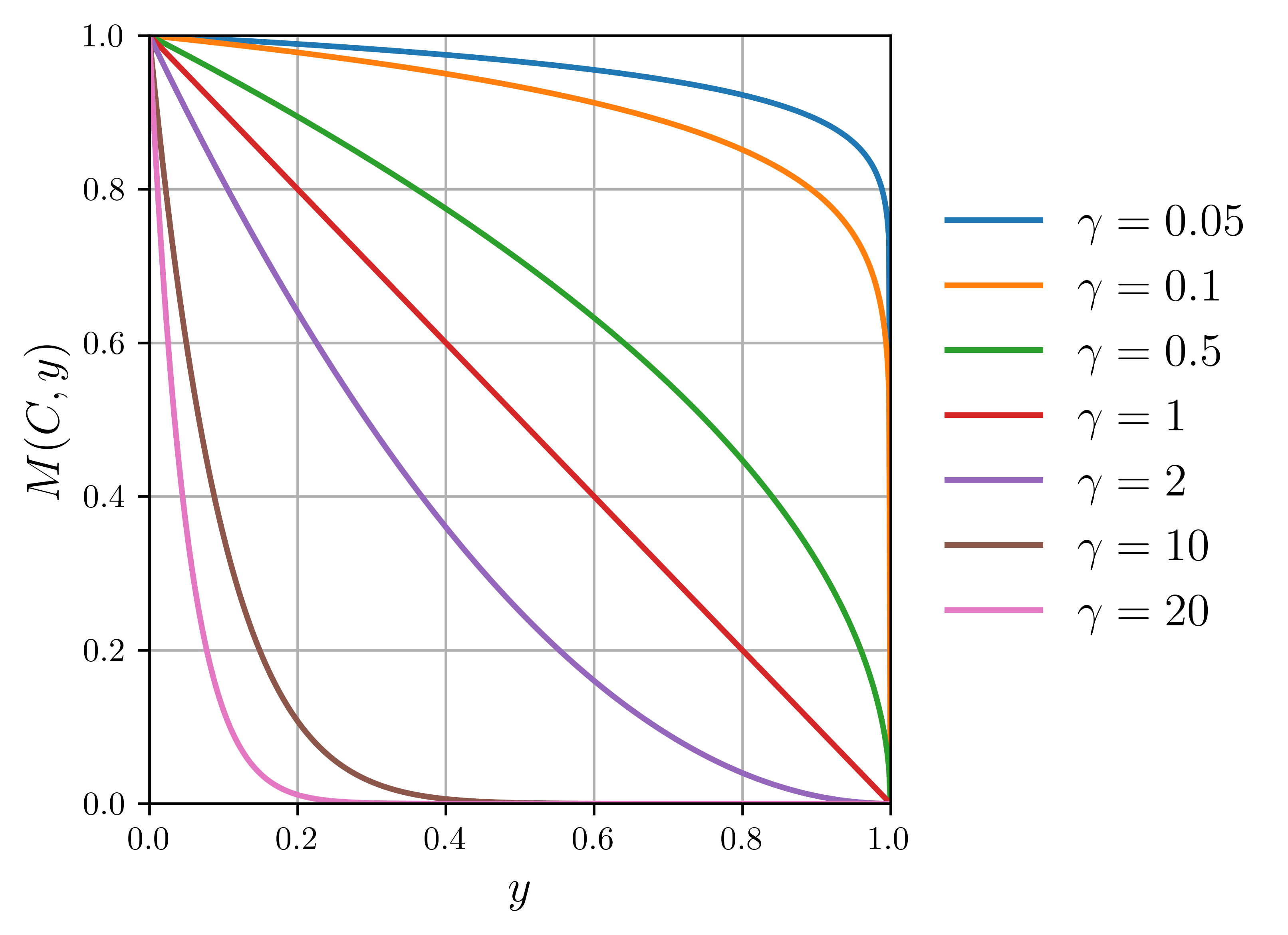}
\caption{\textit{The function \( M(C, y) \) for various values of \( \gamma \), with \( \alpha = 1 \). Low values of \( \gamma \) correspond to highly sensitive cooperators who readily disperse even when few defectors are present. As \( \gamma \) increases, cooperators become more tolerant, leading to reduced migration in defector-rich demes.
 }}
\label{alpha_plot_vs_beta}
\end{figure}

We model migration behavior as both strategy- and context-dependent. Specifically, cooperators exhibit an additional migration probability given by \( M(C, y) = \alpha(1 - y)^{\gamma} \), where \( \alpha > 0 \) governs the intensity of the additional dispersal and \( \gamma > 0 \) determines sensitivity to defectors. In contrast, defectors migrate with a constant probability \( m \) that does not depend on social context, that is to say \( M(D, y) = 0 \). This asymmetry reflects biologically plausible behavior. Cooperators are more inclined to leave unfavorable environments, a pattern consistent with risk avoidance and reciprocity-seeking documented in both social and microbial systems~\cite{PS2002, A2011}. Small values of \( \gamma \) correspond to high sensitivity: cooperators disperse even when only a few defectors are present. Conversely, large \( \gamma \) implies greater tolerance, with dispersal triggered only under widespread defection. This framework captures a spectrum of natural dispersal strategies and introduces a nonlinear migration rule that significantly shapes spatial structure and evolutionary outcomes. Figure \ref{alpha_plot_vs_beta} illustrates these remarks.

Note that the dispersal correction term in the drift function simplifies as
\begin{equation}\label{sec2-eq6}
\begin{split}
&\int_{0}^{1}(y-x)\left(yM(C,y)+(1-y)M(D,y)\right)g(y)dy\\
&=\alpha\int_{0}^1(y-x)\frac{y^{\nu x}(1-y)^{\gamma+\nu(1-x)-1}}{B\left(\nu x,\nu(1-x)\right)}dy\\
&=\alpha\frac{B(\nu x+2,\nu(1-x)+\gamma)-xB(\nu x+1,\nu(1-x)+\gamma)}{B(\nu x,\nu(1-x))}\\
&=\alpha\frac{1-(\gamma+1)x}{\nu+\gamma+1}\frac{\nu x(1-x)}{\nu+1}\frac{B(\nu x+1,\nu(1-x)+\gamma)}{B(\nu x+1,\nu(1-x)+1)}.
\end{split}
\end{equation}
Inserting this into the drift-to-diffusion ratio yields
\begin{align}\label{sec2-eq7}
&\frac{2\mu(x)}{\sigma^2(x)}\nonumber\\
=&\delta\sum_{n=0}^{k}\binom{k}{n}\frac{B\left(\nu x+n+1,\nu(1-x)+k+1-n\right)}{B\left(\nu x+1,\nu(1-x)+1\right)}(a_n-b_n)\nonumber\\
&+\delta\alpha\frac{1-(\gamma+1)x}{\nu+\gamma+1}\frac{B(\nu x+1,\nu(1-x)+\gamma)}{B(\nu x+1,\nu(1-x)+1)}.
\end{align}
Using this expression, the approximation in Eq. (\ref{sec1-eq8}) takes the form
\begin{align}\label{sec2-eq8}
   &\rho_C\approx \frac{1}{Nd}+\frac{\delta}{Nd}\Bigg[\sum_{n=0}^{k}\eta_{n,k}(\nu)\left(a_n-b_n\right)+\alpha Q(\nu,\gamma)\Bigg],
\end{align}
where 
\begin{subequations}
\begin{align}
\eta_{n,k}(\nu) &= \binom{k}{n} \int_{0}^{1}(1 - x) \frac{B(\nu x + n + 1, \nu(1 - x) + k + 1 - n)}{B(\nu x + 1, \nu(1 - x) + 1)} dx,\\
Q(\nu,\gamma) &= \int_{0}^1 (1 - x) \frac{1 - (\gamma + 1)x}{\nu + \gamma + 1} \frac{B(\nu x + 1, \nu(1 - x) + \gamma)}{B(\nu x + 1, \nu(1 - x) + 1)} dx.
\end{align}
\end{subequations}

\begin{figure}
\centering
\includegraphics[height=9cm, width=9cm]{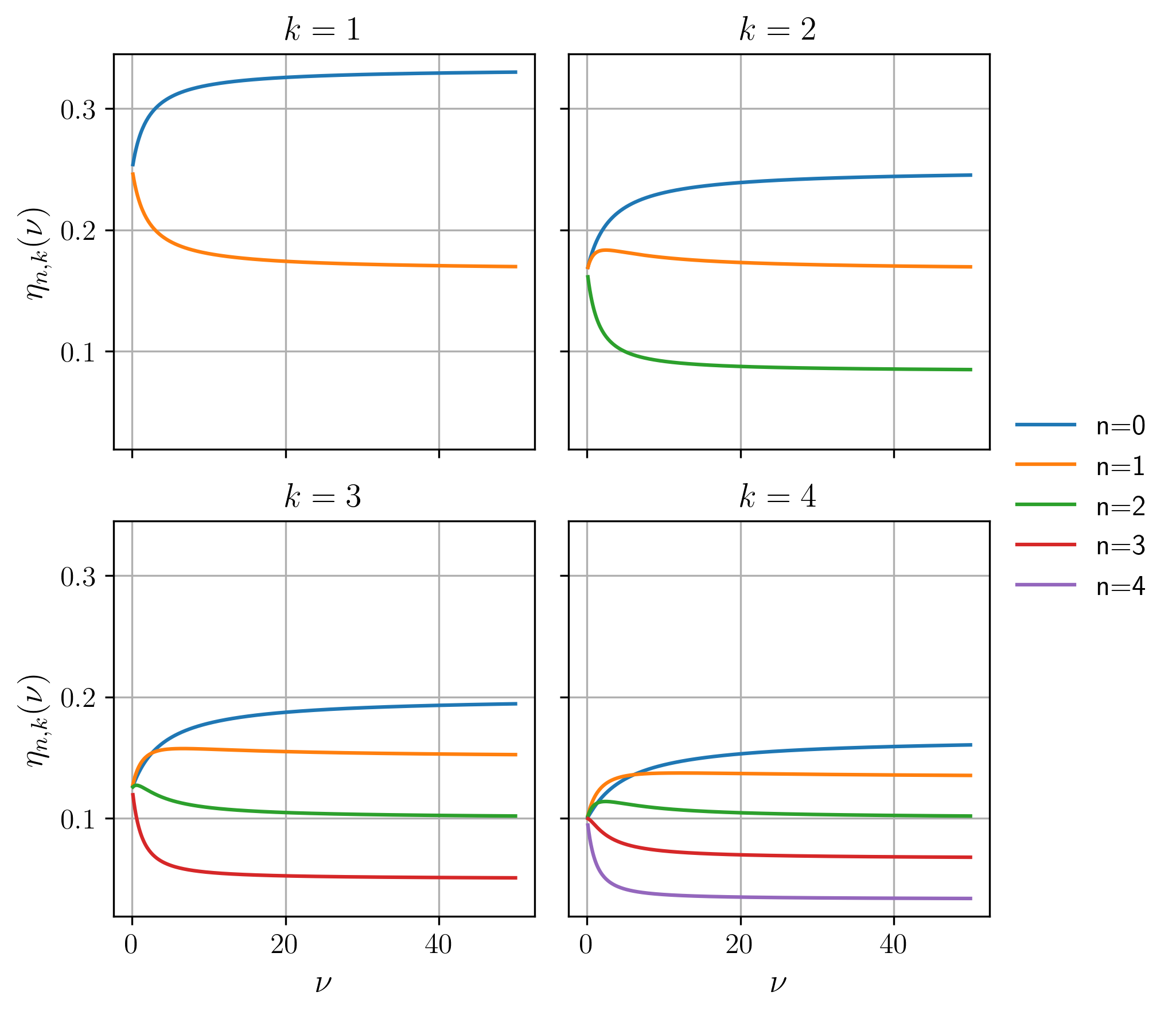}
\caption{\textit{Values of \( \eta_{n,k}(\nu)\) for \( n = 0, 1, \ldots, k \) and group sizes \( k = 1, 2, 3, 4 \). Each \( \eta_{n,k}(\nu) \) represents the weight of to the payoff difference \( a_n - b_n \) in the first-order approximation of $\rho_C$ given by Eq. (\ref{sec2-eq8}). For all \( k \), as \( \nu \to 0 \), these weights converge to \( 1/(2(k+1)) \), reflecting a uniform contribution across group compositions. As \( \nu \) increases, the weights become more skewed, with higher values for low \( n \) (defector-rich groups) and lower values for high \( n \) (cooperator-rich groups), especially for larger \( k \). This illustrates how the baseline migration rate \( \nu \) modulates the relative impact of different interaction compositions on the evolutionary dynamics.}}
\label{coefficients}
\end{figure}

Figure~\ref{coefficients} shows how the baseline migration rate \( \nu \) shapes the coefficients
\((\eta_{n,k}(\nu))_{n=0}^{k}\)
which weight the payoff differences \( a_n - b_n \) in the first-order approximation of the fixation probability \( \rho_C \). These coefficients quantify how group composition and migration affect the selection gradient. Across group sizes \( k = 1,2,3,4 \), the weight associated with \( n = 0 \) (absence of cooperators in a group) increases with \( \nu \), while those for higher \( n \) tend to decrease. This reflects a shift in emphasis toward low-cooperation interactions under high migration, which undermines local assortment and weakens selection for cooperation. As \( \nu \) grows, the coefficients stabilize, suggesting diminishing influence of additional migration. 

\begin{figure}
\includegraphics[height=7cm, width=8.6cm]{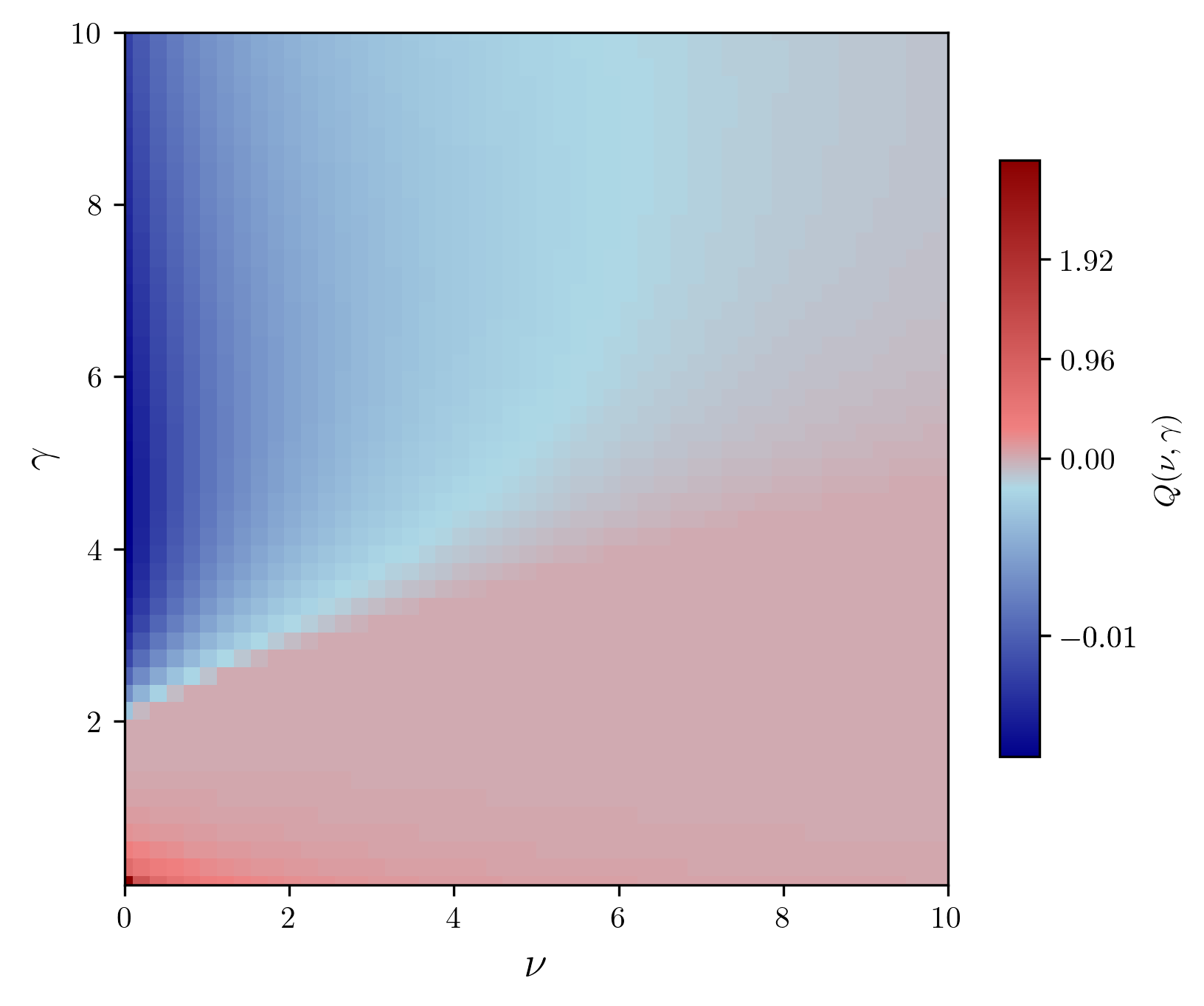}
\caption{\textit{ 
This figure illustrates the function \( Q \), which captures how migration differences between cooperators and defectors influence the fixation probability $\rho_C$, with respect to the parameters \( \nu \) and \( \gamma \). Positive values of \( Q(\nu, \gamma) \) indicate scenarios where conditional migration enhances cooperation by raising $\rho_C$ above its baseline under uniform migration. In contrast, negative values reflect cases where such migration undermines cooperation by lowering $\rho_C$ below its baseline under uniform migration. The blue region corresponds to values where \( Q < 0 \), indicating conditions that work against the fixation of cooperation. Conversely, the red region marks the parameter space where \( Q > 0 \), reflecting scenarios that support cooperative fixation.
}}
\label{sign_of_q}
\end{figure}

In parallel, Figure~\ref{sign_of_q} illustrates how the migration correction term
\(
Q(\nu,\gamma) 
\)
can promote or hinder cooperation depending on the sensitivity parameter \( \gamma \) and the amplification factor \( \alpha \). This term arises from behavioral asymmetry, where cooperators disperse in response to local defector frequency. The heatmap in Figure~\ref{sign_of_q} reveals that \( Q(\nu, \gamma) \) is generally positive when \(\gamma\) is small enough, the threshold increasing as \( \nu \) gets larger, indicating that differential migration favors cooperation. However, for $\gamma$ large, this term becomes negative, showing that indiscriminate dispersal dilutes the benefits of behavioral responsiveness. This underscores how migration strategies that depend on local social context can strongly influence the evolutionary success of cooperation.

In what follows, we examine limiting regimes to interpret these effects biologically.



\subsection{High Migration Regime: $\nu \to \infty$\label{sec2-1}}
Consider the regime where the baseline migration rate is large, i.e., $\nu \to \infty$. In this case, using the asymptotic approximation
\begin{equation}\label{sec2-eqq11}
\Gamma(x + \beta) \approx \Gamma(x) \, x^\beta \quad \text{as } x \to \infty,
\end{equation}
we obtain the approximation
\begin{align}\label{sec2-eq11}
&\quad\frac{B(\nu x + i_1, \nu(1 - x) + j_1)}{B(\nu x + i_2, \nu(1 - x) + j_2)}\nonumber\\
&=\frac{\Gamma(\nu x + i_1)\Gamma(\nu(1-x) + j_1)}{\Gamma(\nu+ i_1+j_1)} \frac{\Gamma(\nu+ i_2+j_2)}{\Gamma(\nu x + i_2)\Gamma(\nu(1-x) + j_2)}\nonumber\\
&\approx \frac{(\nu x)^{i_1}(\nu(1 - x))^{j_1}}{\nu^{i_1+j_1}} \frac{\nu^{i_2+j_2}}{(\nu x)^{i_2}(\nu(1 - x))^{j_2}} \nonumber\\
&\approx x^{i_1-i_2} (1 - x)^{j_1-j_2},
\end{align}
for any $i_1,i_2,j_1,j_2\geq0$, as $\nu\rightarrow\infty$.
This shows the asymptotic estimates
\begin{equation}\label{sec2-eq12}
\begin{split}
\eta_{n,k}(\nu)
&\approx \binom{k}{n}\int_{0}^{1}x^{n} (1 - x)^{k+1-n}dx\\
&\approx \binom{k}{n}B(n+1,k+2-n)
\end{split}
\end{equation}
and
\begin{equation}\label{sec2-eq13}
Q(\nu,\gamma)
 \approx \frac{1}{\nu + \gamma + 1} \int_0^1 \left(1 - (\gamma + 1)x\right)(1 - x)^\gamma dx\approx o(1).
\end{equation}
Inserting Eqs. (\ref{sec2-eq12}) and (\ref{sec2-eq13}) into Eq. (\ref{sec2-eq8}), the fixation probability of type $C$ will be approximated as
\begin{equation}\label{sec2-eq14}
\rho_C \approx \frac{1}{Nd} + \frac{\delta}{Nd} \sum_{n = 0}^{k} \binom{k}{n} B(n+1, k+2 - n)(a_n - b_n).
\end{equation}
This recovers the expression for the fixation probability of type \( C \) in a large well-mixed population (see Eq.~(23) in Kroumi \textit{et al.}~\cite{KML2022}). In the regime of large \( \nu \), both cooperators and defectors exhibit very high mobility, where the additional migration advantage of cooperators vanishes, and the population subdivision has no effect. The dynamics revert to a well-mixed regime, where selection acts solely within demes, and cooperation can only evolve if individually advantageous. In this setting, neither \( \alpha \) nor \( \gamma \) influences evolutionary outcomes.


\subsection{Low Migration Regime: $\nu << 1$\label{sec2-2}}

Next, we examine the opposite regime in which the baseline migration rate is very low, i.e., \( \nu \to 0 \). In this limit, the effect of the structure on the evolutionary dynamics becomes maximal. In this case, we have the approximations
\begin{align}\label{sec2-eq15}
\eta_{n,k}(\nu)
&\approx \binom{k}{n}\int_{0}^{1}(1 - x)\frac{B\left(n+1,k+1-n\right)}{B\left(1,1\right)}dx\nonumber\\
&\approx\frac{1}{2(k+1)}
\end{align}
and
\begin{align}
Q(\nu,\gamma)
&\approx \int_0^1 \frac{1 - (\gamma + 1)x}{\gamma + 1} (1 - x) B(1, \gamma) dx\nonumber \\
&\approx \int_0^1 \frac{1 - (\gamma + 1)x}{\gamma + 1} (1 - x) \frac{1}{\gamma} dx\nonumber \\
&\approx \frac{2 - \gamma}{6\gamma(\gamma + 1)}.
\end{align}

Inserting these approximations in Eq. (\ref{sec2-eq8}), the fixation probability of $C$ can be written as
\begin{align}\label{sec2-eq18}
\rho_C &\approx \frac{1}{Nd} + \frac{\delta}{Nd} \left( \sum_{n=0}^{k} \frac{1}{2(k+1)} (a_n - b_n)+ \alpha \cdot \frac{2 - \gamma}{6\gamma(\gamma + 1)} \right).
\end{align}
This expression reveals a nuanced effect of the parameters \( \alpha \) and $\gamma$ on $\rho_C$. We distinguish two main cases:

\begin{itemize}
    \item If \( 0 < \gamma < 2 \), the additional migration term increases the fixation probability of \( C \), with higher values of \( \alpha \) further enhancing \( \rho_C \). In this regime, cooperators are highly sensitive to defection and disperse early, even at low defector frequencies. This proactive migration facilitates their escape from unfavorable environments and promotes clustering in cooperative demes, thereby boosting their relative fitness and favoring the evolution of cooperation.

    \item Conversely, if \( \gamma > 2 \), the additional migration term becomes detrimental, with higher values of \( \alpha \) reducing \( \rho_C \). Cooperators exhibit low sensitivity to defection and delay migration until defector frequency is high. This reactive behavior limits early escape from exploitative groups, weakening assortment and hindering the formation of cooperative clusters, thereby reducing the likelihood of cooperative fixation.
\end{itemize}

Finally, when \( \gamma \) is very large, cooperators become highly tolerant of defectors and rarely migrate in response. As a result, the additional migration term has negligible effect, the selection pressure from differential migration vanishes, and the fixation probability \( \rho_C \) becomes effectively independent of \( \alpha \), given by
\begin{align}
\rho_C &\approx \frac{1}{Nd} + \frac{\delta}{Nd} \sum_{n=0}^{k} \frac{1}{2(k+1)} (a_n - b_n).
\end{align}


\subsection{High sensitivity regime: \( \gamma << 1\)\label{sec2-3}}

Now, let us examine the scenario in which cooperators exhibit extreme sensitivity to the presence of defectors, corresponding to the limit \( \gamma << 1 \) for a fixed $\nu>0$. In this regime, the additional migration term \( M(C, y) = \alpha(1 - y)^\gamma \) approaches $1$ for any $y<1$: even a small proportion of defectors in a deme causes cooperators to disperse with near-maximal probability. Note the asymptotic approximation
\begin{align*}
Q(\nu,\gamma)\approx \int_0^1 \frac{1 - x}{\nu} dx = \frac{1}{2\nu},
\end{align*}
yielding the following estimate 
\begin{align}\label{sec2-eq20}
&\rho_C \approx \frac{1}{Nd} + \frac{\delta}{Nd} \left[ \sum_{n=0}^k \eta_{n,k}(\nu)(a_n - b_n) + \frac{\alpha}{2\nu} \right].
\end{align}

This result highlights a key feature of high-sensitivity dispersal: cooperation is most strongly promoted when $\nu<<1$, as cooperators quickly leave mixed groups due to strong aversion to defectors. This facilitates rapid clustering into cooperative demes, enhancing positive assortment and favoring cooperation, as supported by both simulations and microbial experiments \cite{PS2002, NXF2010, A2011}. As the baseline migration rate $\nu$ increases, defectors also become more mobile, undermining assortment by infiltrating cooperative groups and reducing overall cooperation. Increasing \( \alpha \) further boosts the fixation probability of cooperators by amplifying this selective dispersal. These findings support the broader view that behavioral responsiveness can substitute for kin structure or fixed spatial constraints. They provide analytical backing for the idea that "walking away" is most effective under limited mobility and moderate connectivity \cite{FZHTLG2022, RBR2023}, and extend previous work by explicitly quantifying how sensitivity and migration interact to shape evolutionary outcomes.


\subsection{Low sensitivity regime: \( \gamma >> 1\)\label{sec2-4}}
We now consider the opposite behavioral extreme: cooperators who exhibit very low sensitivity to the presence of defectors, corresponding to the limit \( \gamma \gg 1 \). In this regime, cooperators tolerate high frequencies of defectors in their local environment before migrating. Using the approximation in Eq.~(\ref{sec2-eqq11}), we find that 
\begin{equation}\label{sec2-eq21}
    \begin{split}
Q(\nu,\gamma)&\approx \int_0^1 -x(1-x)\frac{\gamma^{\nu(1-x)}\Gamma(\nu+2)}{\Gamma(\nu(1-x)+1)\gamma^{\nu+1}} dx \\
&\approx o(\gamma^{-1}),
    \end{split}
\end{equation}
implying that the additional migration term for cooperators vanishes asymptotically. Consequently, the fixation probability of a single cooperator reduces to
\begin{equation}\label{sec2-eq22}
\rho_C \approx \frac{1}{Nd} + \frac{\delta}{Nd}\sum_{n=0}^k \eta_{n,k}(\nu) (a_n - b_n) .
\end{equation}
This result holds for any values of \( \nu \) and \( \alpha \). Biologically, it reflects a scenario in which cooperators remain in their natal groups regardless of social conditions, effectively behaving like defectors in terms of dispersal. As a result, the advantage of behavioral plasticity is nullified, and evolutionary outcomes are determined solely by the payoff regime and population structure.

Our finding aligns with simulation-based work suggesting that overly tolerant strategies suppress cooperative assortment \cite{FZHTLG2022}. It also complements recent models of tolerance-based migration on networks \cite{PABS2023}, where high tolerance thresholds reduce the likelihood of cooperative clusters. In our model, this effect emerges analytically as the loss of sensitivity reduces the migration asymmetry between cooperators and defectors, collapsing the multi-level selection gradient. These insights reinforce a recurring theme in cooperation theory: mobility must be both conditional on and responsive to social context to promote cooperation effectively. When movement becomes insensitive, structure alone is insufficient to maintain cooperation  \cite{R2007,MAB2024}.

\section{Public goods game\label{sec3}}
To better illustrate the impact of the model parameters on the fixation of cooperation, we consider a linear public goods game by groups of size \( k+1 \). In each group, a cooperator contributes a benefit \( b \) to the group at a personal cost \( c \), while defectors contribute nothing and incur no cost. The total benefit is shared equally among all group members, yielding the following payoffs
\begin{align}
a_n &= \frac{n+1}{k+1}b - c,\\
b_n& = \frac{n}{k+1}b,
\end{align}
for a cooperator and a defector, respectively. The resulting payoff difference is
\[
a_n - b_n = \frac{b}{k+1} - c,
\]
which does not depend on \( n \).

We plug these payoffs into Eq. (\ref{sec2-eq8}), and using
\begin{align}\label{sec3-eq3}
& \sum_{n=0}^{k}\eta_{n,k}(\nu)=\int_{0}^{1}(1-x)dx=\frac{1}{2},
\end{align}
 the fixation probability of type $C$ can be approximated as 
\begin{equation}\label{sec3-eq4}
    \rho_C\approx \frac{1}{Nd}+\frac{\delta}{2Nd}\left( \frac{b}{k+1} -c+2\alpha Q(\nu,\gamma) \right).
\end{equation}

We say that selection favors the fixation of cooperation if \( \rho_C > 1/(Nd) \), where \( 1/(Nd) \) is the expected fixation probability under neutrality—that is, in the case of equal payoffs and assuming uniform migration governed solely by the parameter \( m \). A remark concerns the interpretation of the condition under which selection favors the fixation of type \( C \). This condition can be equivalently written as
\begin{equation}\label{sec3-eq16}
\frac{b}{k+1} + 2\alpha Q(\nu,\gamma) > c.
\end{equation}
This can be viewed as a generalized form of Hamilton’s rule \cite{H1964}, where cooperation evolves if the combined benefits outweigh the cost. The term \( \frac{b}{k+1} \) represents the direct benefit scaled by within-group relatedness, while \( 2\alpha Q(\nu,\gamma) \) captures an indirect benefit arising from migration-driven assortment. This second term reflects the advantage cooperators gain by relocating away from defectors and forming more cooperative clusters. Together, they define an effective relatedness-adjusted benefit, illustrating how spatial structure and behavioral plasticity in migration can promote the evolution of cooperation. 

Now, when migration is uniform ($\alpha = 0$ or $\gamma>>1$) or when the baseline migration rate is high ($\nu>>1$), the condition for selection to favor the fixation of cooperators simplifies to
\[
\frac{b}{c} > k + 1,
\]
and does not depend on the baseline migration rate \( \nu \). This matches the classical condition for selection to favor the fixation of cooperation in a well-mixed population \cite{VN2012}, indicating that population structure alone does not promote cooperation in this setting. Notably, as the group size \( k+1 \) increases, this threshold becomes increasingly difficult to overcome. For large \( k \), even high benefit \( b \) fails to offset the cost \( c \), and selection consistently disfavors the fixation of cooperation. 

In the remainder of this section, we incorporate an additional migration term for cooperators and assume a finite baseline migration rate. We will analyze the fixation probability \( \rho_C \) and derive conditions on \( b/c \) under which selection favors the fixation of cooperation across different parameter regimes. We begin with the low migration regime (\( \nu \ll 1 \)), showing that both \( \alpha \) and \( \gamma \) positively influence the evolution of cooperation. Next, we examine a case of high sensitivity  (\( \gamma = 1 \)), where cooperation is most favored under low \( \nu \) or large \( \alpha \). Finally, we consider a case of moderate sensitivity (\( \gamma = 3 \)) and find that cooperation is best promoted at intermediate \( \nu \) vaules or high \( \alpha \) values.


\subsection{Low baseline migration rate\label{sec3-1}}
Suppose the baseline migration rate is low, i.e., \( \nu \ll 1 \). In this regime, the fixation probability of cooperators simplifies to
\begin{align*}
\rho_C &\approx \frac{1}{Nd} + \frac{\delta}{2Nd} \left( \frac{b}{k+1} - c + \alpha \cdot \frac{2 - \gamma}{3\gamma(\gamma + 1)} \right).
\end{align*}
As detailed in Section~\ref{sec2-2}, the effect of the correction term depends critically on the sensitivity parameter \( \gamma \). When \( \gamma < 2 \), the additional migration term is positive, leading to a fixation probability \( \rho_C \) that exceeds its neutral baseline. This implies that in highly sensitive regimes, conditional dispersal enhances cooperation: cooperators are more likely to exit defector-dominated demes early, thereby facilitating the formation of cooperative clusters. In contrast, when \( \gamma > 2 \), the migration response becomes counterproductive. Although cooperators still disperse, they do so neither early enough to form new clusters nor late enough to stabilize existing ones. In this case, migration disrupts group structure without providing a cooperative advantage, resulting in a decreased fixation probability.

Moreover, the influence of \( \gamma \) on \( \rho_C \) is non-monotonic. For values of \( \gamma \) in the interval \( (0, 2 + \sqrt{6}) \), increasing sensitivity promotes cooperation by increasing \( \rho_C \). However, beyond the critical threshold \( \gamma = 2 + \sqrt{6} \), further increases in sensitivity become detrimental, reducing the likelihood of cooperative fixation. Therefore, the most favorable conditions for the evolution of cooperation occur when \( \gamma \ll 1 \), corresponding to strong aversion to defectors. The least favorable scenario arises near the critical point \( \gamma = 2 + \sqrt{6} \), where migration remains active but fails to support cooperative dynamics.

\begin{figure}
\centering
\includegraphics[height=7cm, width=9cm]{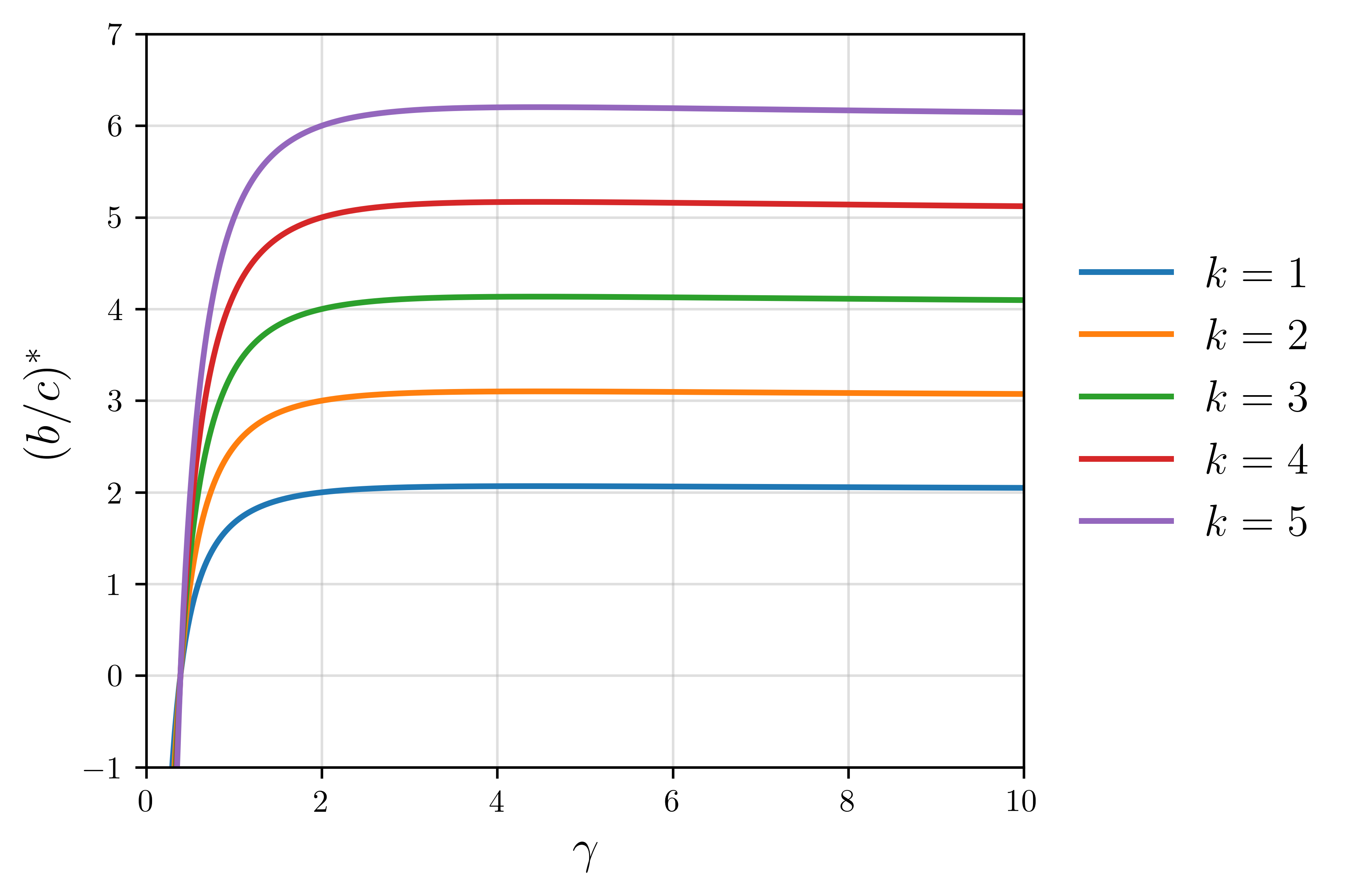}
\caption{\textit{This figure depicts the critical threshold \( (b/c)^* \) as a function of the sensitivity parameter \( \gamma > 0 \), assuming \( \alpha/c = 1 \), for different group sizes \( k = 1, 2, 3, 4, 5 \), under the assumption that \( \nu \ll 1 \). When \( \gamma < \gamma^* \), corresponding to high sensitivity, \( (b/c)^* \leq 0 \) and fixation of cooperation is favored. For \( \gamma^* < \gamma < 2 \), the threshold satisfies \( 0 < (b/c)^* < k + 1 \), indicating a less stringent condition than in a well-mixed population. In contrast, when \( \gamma > 2 \), the threshold exceeds \( k + 1 \), making cooperation harder to fix.
}}
\label{threshold_public_low}
\end{figure}

Now, we are interested on the condition for selection to favor the fixation of type $C$, that is to say $\rho_C>1/(Nd)$.
In this regime, the condition will take the form
\begin{equation}\label{sec3-eq7}
\frac{b}{c} > \left(\frac{b}{c}\right)^* = (k+1)\left(1 - \frac{\alpha}{c} \cdot \frac{2 - \gamma}{3\gamma(\gamma + 1)}\right).
\end{equation}
Solving the equation
\[
1 - \frac{\alpha}{c} \cdot \frac{2 - \gamma}{3\gamma(\gamma + 1)} = 0
\]
yields the critical value
\[
0 < \gamma^* = \frac{\sqrt{\left(\frac{\alpha}{c} + 3\right)^2 + 21\frac{\alpha}{c}} - \frac{\alpha}{c} - 3}{6} < 2,
\]
which marks the transition beyond which we have \( \left(\frac{b}{c}\right)^* \geq 0 \). For all \( \gamma \leq \gamma^* \), selection will favor the fixation of cooperation for any \( b > 0 \), \( c > 0 \), and any group size \( k \geq 1 \).
As \( \gamma \) increases beyond \( \gamma^* \), the threshold satisfies \( (b/c)^*>0 \). When \( \gamma^* < \gamma < 2 \), the condition for selection to favor cooperation remains less stringent than in a well-mixed population, since \( 0 < (b/c)^* < k + 1 \). However, once \( \gamma > 2 \), we find \( (b/c)^* > k + 1 \), implying that the requirement for cooperation to evolve becomes stricter than in a well-mixed setting. Finally, in the limit of very low sensitivity (i.e., \( \gamma \to \infty \)), the threshold converges to \( (b/c)^* = k + 1 \), precisely matching the classical result for well-mixed populations.
These analytical insights are illustrated in Figure~\ref{threshold_public_low}, which shows how the critical benefit-to-cost threshold varies with sensitivity and group size.

\subsection{Linear sensitivity : $\gamma=1$\label{sec3-2}}

Consider the case of linear sensitivity, i.e., $\gamma=1$, where the additional migration behavior of cooperators is given by \( M(C, y) = \alpha(1 - y) \). In this scenario, a cooperator's tendency to migrate decreases linearly with the local frequency of cooperators in its deme. The corresponding migration-related correction in the fixation probability simplifies to
\begin{align}\label{sec3-eq11}
Q(\nu, 1) = \int_{0}^{1} (1 - x)\frac{1 - 2x}{\nu + 2} \, dx = \frac{1}{6(\nu + 2)}.
\end{align}
This leads to the following approximation 
\begin{align}\label{sec3-eq12}
\rho_C \approx \frac{1}{Nd} + \frac{\delta}{2Nd} \left( \frac{b}{k+1} - c + \frac{\alpha}{3(\nu + 2)} \right).
\end{align}

Two key insights emerge from this expression:
\begin{itemize}
    \item \textbf{Effect of Baseline Migration Rate (\( \nu \))}: As \( \nu \) increases, the migration correction term \( \frac{\alpha}{3(\nu + 2)} \) decreases, reducing \( \rho_C \). This reflects the erosion of spatial structure at higher migration rates, which weakens assortment and makes cooperation harder to fix.
    
    \item \textbf{Effect of Sensitivity Scaling (\( \alpha \))}: Increasing \( \alpha \) amplifies the migration correction, thereby promoting cooperative fixation. Higher values of \( \alpha \) strengthen the response of cooperators to local defection, encouraging them to relocate to more cooperative environments, as observed in models of adaptive migration \cite{YW2011,ISSW2013}.
\end{itemize}

We now examine the critical threshold \( \left( \frac{b}{c} \right)^* \) that the benefit-to-cost ratio must exceed for selection to favor the fixation of cooperation, i.e., $\rho_C>1/(Nd)$. From Eq.~\eqref{sec3-eq12}, this threshold is given by
\[
\left( \frac{b}{c} \right)^* = (k+1)\left(1 - \frac{\alpha}{c} \cdot \frac{1}{3(\nu + 2)}\right).
\]
This threshold is always below \( k + 1 \), the corresponding value in a well-mixed population, indicating that structured populations can promote cooperation more easily. Moreover, \( \left( \frac{b}{c} \right)^* \) increases with respect to \( \nu \), ranging from
\[
\left( \frac{b}{c} \right)^* \Big|_{\nu \to 0^+} = (k+1)\left(1 - \frac{\alpha}{6c} \right)
\quad \text{to} \quad
\left( \frac{b}{c} \right)^* \Big|_{\nu \to \infty} = k + 1.
\]
This reflects the fact that cooperation is most favored when the baseline migration rate is low. In this regime, cooperators benefit from strong mobility due to the additional migration term, while defectors remain largely stationary, allowing cooperators to assort effectively. As \( \nu \) increases, defectors gain more mobility, dispersal becomes less selective, and the advantage of assortment diminishes. Consequently, the population approaches a well-mixed state, and a higher benefit-to-cost ratio is required for cooperation to fix.


\subsection{Moderate sensitivity case: $\gamma=3$\label{sec3-3}}
The last scenario to be considered is a case of moderate sensitivity for \( \gamma = 3 \). In this setting, the additional migration term for cooperators responds weakly to the presence of defectors, reflecting a more tolerant behavioral strategy. The migration-related correction term simplifies as
\begin{align}\label{sec3-eq17}
Q(\nu,3)\nonumber &= \int_{0}^{1}(1-x) \frac{1-4x}{\nu+4} \cdot \frac{B(\nu x+1,\nu(1-x)+3)}{B(\nu x+1,\nu(1-x)+1)}dx \nonumber\\
&= \int_{0}^{1}(1-x) \frac{1-4x}{\nu+4} \cdot \frac{(\nu(1-x)+2)(\nu(1-x)+1)}{(\nu+3)(\nu+2)}dx \nonumber\\
&= \frac{3\nu^{2} - 20}{60(\nu+2)(\nu+3)(\nu+4)}.
\end{align}
Consequently, the fixation probability of a single cooperator is approximated by
\begin{equation}\label{sec3-eq18}
\rho_C \approx \frac{1}{Nd} + \frac{\delta}{2Nd} \left( \frac{b}{k+1} - c + \alpha \cdot \frac{3\nu^{2} - 20}{30(\nu+2)(\nu+3)(\nu+4)} \right).
\end{equation}

\begin{figure}
\centering
\includegraphics[height=6cm, width=8cm]{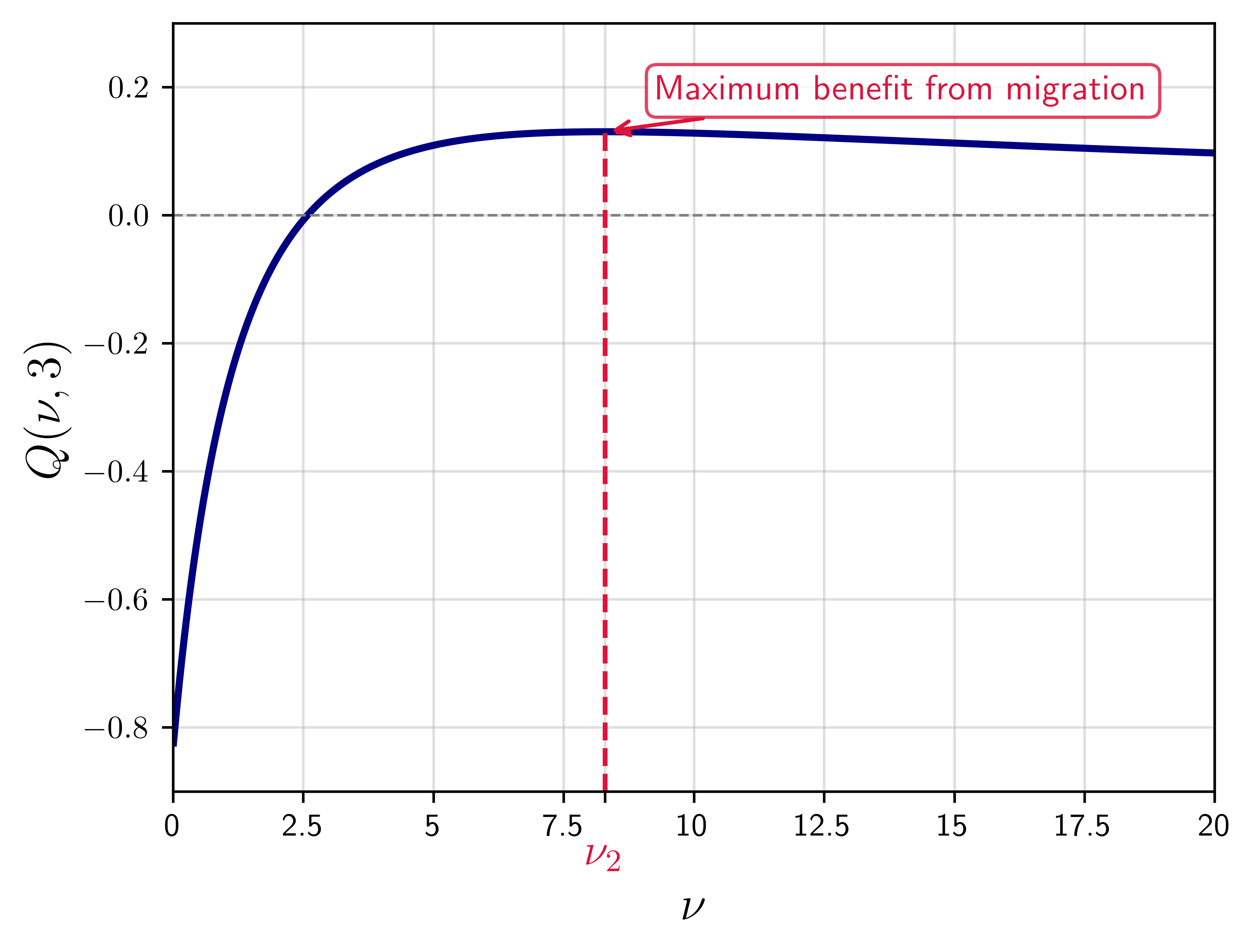}
\caption{\textit{This curve depicts the function \( Q(\nu,3) \) as a function of the baseline migration rate \( \nu \). The curve is initially negative, becomes positive, and reaches its maximum at the critical point \( \nu_2 \approx 8.29 \). This value marks the most favorable scenario for the fixation of cooperation, where moderate migration maximizes the benefits of context-dependent dispersal. For \( \nu > \nu_2 \), the benefit of migration declines as dispersal becomes less selective, reducing the ability of cooperators to form and maintain cooperative clusters.}}
\label{threshold_publi_sensitivity_low}
\end{figure}

In contrast to the case \( \gamma = 1 \), where increasing \( \alpha \) or decreasing \( \nu \) always promotes the fixation of cooperation, the effects here are more complex. For \( \nu < \nu_1 = \sqrt{20/3} \approx 2.58 \), the correction term is negative, and increasing \( \alpha \) reduces \( \rho_C \). Fixation of cooperation is least favored when \( \nu \ll 1 \), as cooperators lack the mobility needed to escape partially defective demes. As \( \nu \) increases toward \( \nu_1 \), the fixation probability of $C$ improves due to greater dispersal opportunity. Beyond \( \nu_1 \), \( Q(\nu,3) \) becomes positive and migration starts to favor $C$ fixation. A critical point \( \nu_2 \approx 8.29 \) maximizes this benefit, defined by
\[
\left. \frac{d}{d\nu} Q(\nu, 3) \right|_{\nu = \nu_2} = 0.
\]
For \( \nu > \nu_2 \), further increases in migration reduce \( \rho_C \), as high mobility neutralizes behavioral differences, and the dynamics approach the well-mixed case. These trends are visualized in Figure~\ref{threshold_publi_sensitivity_low}.

Biologically, this situation reflects the trade-off between responsiveness and mobility. Moderately sensitive cooperators do not react strongly to few defectors and may remain in suboptimal demes. When migration is too limited, this traps them in exploitative environments. Moderate migration rates enable relocation to cooperative clusters and improve evolutionary outcomes. However, at high baseline migration rate, dispersal becomes indiscriminate, and the benefits of conditional movement are lost.

We now analyze the threshold \( \left(\frac{b}{c}\right)^* \) and compare it to the well-mixed benchmark \( k + 1 \). From Eq.~\eqref{sec3-eq18}, this threshold is
\begin{equation}\label{sec3-eq20}
\left(\frac{b}{c}\right)^* = (k+1)\left(1 + \frac{\alpha}{c} \cdot \frac{20 - 3\nu^{2}}{30(\nu+2)(\nu+3)(\nu+4)}\right),
\end{equation}
which varies non-monotonically with \( \nu \). Initially, it decreases with increasing \( \nu \), starting from
\begin{equation}\label{sec3-eq21}
\left(\frac{b}{c}\right)^* \Big|_{\nu \to 0^+} = (k+1)\left(1 + \frac{\alpha}{36c}\right) > k + 1,
\end{equation}
and reaching a minimum at \( \nu = \nu_2 \), where
\begin{equation}\label{sec3-eq22}
\left(\frac{b}{c}\right)^* \Big|_{\nu = \nu_2} = (k+1)\left(1 - \frac{\alpha}{c} \cdot Q(\nu_2, 3)\right) < k + 1.
\end{equation}
For \( \nu > \nu_2 \), the threshold rises and approaches the well-mixed value
\begin{equation}\label{sec3-eq23}
\left(\frac{b}{c}\right)^* \Big|_{\nu \to \infty} = k + 1.
\end{equation}


\section{Discussion}

In this paper, we have examined the role of conditional dispersal in the evolution of cooperation within structured populations. By introducing a migration rule that depends on local social composition, we extend classical models to account for behavioral plasticity. We show that a cooperator’s propensity to migrate modulated by the function \( M(C,y) = \alpha(1 - y)^\gamma \), where $y$ is the frequency of cooperators in the focal deme, significantly influences fixation probabilities. In the context of a finite-island Moran model with a large number $d$ of demes each of size $N$, this migration behavior enables cooperators to leave unfavorable demes and form more favorable clusters, fostering dynamic positive assortment. Such assortment is essential for overcoming the inherent disadvantage of cooperation in well-mixed populations \cite{H1964, R2004}, and aligns with the principles of multi-level selection: while defectors outperform cooperators within groups, groups dominated by cooperators outperform others at the metapopulation scale \cite{W1975}.

This framework reflects a continuum of dispersal strategies observed in nature. Low \( \gamma \) models proactive strategies found in social insects and mobile foragers, who abandon suboptimal environments  \cite{DGVW2004, BSR2001}. It also captures microbial behaviors such as chemotaxis and starvation-induced aggregation \cite{PWYLDSSA2003, SZQ2000}, where cells move in response to local social or environmental cues. High \( \gamma \), by contrast, models tolerant behaviors seen in biofilm-forming bacteria, sessile marine invertebrates, and clonal plant systems, where dispersal is suppressed even under unfavorable conditions \cite{CSG1999, O1986, HKMR1999}. Our results emphasize that the evolutionary outcome of cooperation depends not just on the presence of migration, but on how dispersal decisions respond to local context.

A key insight of our model is that the evolutionary advantage of differential migration hinges on the relationship between the baseline migration rate \( \nu=Nm \) ($m$ is the baseline migration probability shared by all individuals)  and the behavioral sensitivity \( \gamma \). For cooperation to invade and fix, cooperators must exhibit higher mobility than defectors, allowing them to escape exploitative groups and cluster with other cooperators. When sensitivity is high (i.e., \( \gamma \ll 1 \)), even small amounts of defection prompt cooperators to disperse. In this regime, low baseline migration rates are most favorable, as they maintain a strong contrast in mobility between cooperators and defectors: cooperators relocate proactively, while defectors remain largely immobile. However, as \( \nu \) increases, defectors gain greater mobility, eroding this asymmetry. The resulting homogenization undermines cooperative clustering and diminishes the benefits of conditional dispersal.

In contrast, when sensitivity is moderate (e.g., \( \gamma = 3 \)), cooperators are less responsive to low defector frequencies and may remain in partially defective demes. Under a low baseline migration rate $\nu$, this reluctance to move hinders their ability to escape and aggregate with other cooperators, reducing fixation probability. As \( \nu \) increases to a moderate level, cooperators gain sufficient mobility to relocate to more favorable environments, improving cooperation. However, excessive migration again undermines this effect, as both cooperators and defectors disperse freely, neutralizing the impact of sensitivity and preventing stable cooperative clusters.

Our findings complement recent studies demonstrating that spatial structure alone does not guarantee the emergence of cooperation to evolve. Moawad \textit{et al.} \cite{MAB2024} show that cooperation fails to evolve in structured demes unless migration is linked to local conditions. Similarly, Su \textit{et al.} \cite{SMP2023} find that cooperation can emerge in dynamic networks through temporal restructuring, even when each static configuration disfavors it. These studies echo our conclusion that migration must be context-responsive to support cooperation.

A central contribution of our work is the derivation of an analytical expression for the fixation probability of a single cooperator that explicitly incorporates migration asymmetry. When cooperators are highly sensitive to defection (\( \gamma \to 0 \)), the fixation probability increases sharply—even in settings where such an increase would fail under standard conditions. This aligns with prior simulations of "walk-away" strategies \cite{A2011, PS2002}, but our model is the first to offer a continuous-time, diffusion-based analysis across a range of sensitivities and migration rates. Importantly, the efficacy of differential dispersal depends critically on \( \nu \): it is maximized under low values of $\nu$ and vanishes as $\nu$ becomes large \cite{WT2004, CA2016}.

We further identify a critical threshold for the ratio benefit-to-cost required for the fixation of cooperation to be favored and examine how it scales with group size, migration rate, and sensitivity. This threshold increases with \( \gamma \), reflecting that tolerant cooperators delay beneficial dispersal and fail to generate sufficient assortment. However, if cooperators are too reactive (very low \( \gamma \)), premature dispersal can fragment emerging clusters. These findings predict an optimal intermediate sensitivity, a conclusion supported by recent theoretical work. For example, Fahimipour \textit{et al.} \cite{FZHTLG2022} show that moderate avoidance stabilizes cooperative "safe havens" on networks, while extreme sensitivity leads to instability. Pattni \textit{et al.} \cite{PABS2023} similarly find that overly low tolerance disrupts the formation of cooperative groups in evolving networks. Our model generalizes these insights within the Moran-diffusion framework and extends them to multi-player settings.

Overall, this work bridges key themes in cooperation theory: group selection, spatial dynamics, and behavioral plasticity. It supports the idea that adaptive dispersal can substitute for kin-based or static structural mechanisms \cite{NM1992, R2007}. By linking local migration decisions to global fixation outcomes through analytical derivations \cite{EN1980, E2004}, we show that population structure alone is insufficient and that interaction patterns need to adapt to context as well. In structured populations, cooperation thrives when cooperators can find each other, and context-dependent movement may be among the most effective ways to ensure they do.

\begin{widetext}
\appendix
\section{\label{appA}Approximation by a diffusion process}
To analyze the evolutionary dynamics of cooperation in structured populations, we employ a diffusion approximation based on a two-timescale separation framework. This method is particularly suited for systems where a slow variable, such as the global frequency of cooperators \( X(t) \), evolves over generations, while fast variables, such as the distribution of deme types \( \mathbf{Y}(t) \), rapidly converge to a quasi-equilibrium distribution \( \mathbf{v}(X(t)) \) conditional on \( X(t) \). The resulting deviation process \( \mathbf{Y}(t) = \mathbf{X}(t) - \mathbf{v}(X(t)) \) captures short-term fluctuations around this equilibrium. This approximation technique has been rigorously developed in the work of Ethier and Nagylaki~\cite{EN1980} and applied to various models of population structure. In this study, we verify the required conditions in the context of a variant of the island model introduced by Lessard~\cite{L2009} and further refined by Kroumi and Lessard~\cite{KL2025}. Our analysis demonstrates that a suitably rescaled \(X(t)\) converges in distribution to a one-dimensional diffusion that averages reproduction, selection, and migration, while the fast component \(\mathbf Y(t)\) converges to zero in probability.

We begin by computing the first, second, and fourth conditional moments of the one–step change in the metapopulation frequency of cooperators,
\(\Delta X(t)=X(t+1)-X(t)\).  Conditioning on the deme–type vector
\(\mathbf Z(t)=\mathbf z\), we obtain 
\begin{subequations}\label{AppendixAA}
\begin{align}
\mathbb{E}\!\left[\Delta X(t)\,\big|\,\mathbf Z(t)=\mathbf z\right]
  &=\frac{\mu(\mathbf z)}{(Nd)^{2}}+o\!\bigl((Nd)^{-2}\bigr),\\[2pt]
\mathbb{E}\!\left[\Delta X^{2}(t)\,\big|\,\mathbf Z(t)=\mathbf z\right]
  &=\frac{\sigma^{2}(\mathbf z)}{(Nd)^{2}}+o\!\bigl((Nd)^{-2}\bigr),\\[2pt]
\mathbb{E}\!\left[\Delta X^{4}(t)\,\big|\,\mathbf Z(t)=\mathbf z\right]
  &=o\!\bigl((Nd)^{-2}\bigr),
\end{align}
\end{subequations}
where
\begin{subequations}\label{A-eq5}
\begin{align}
\mu(\mathbf{z}) &= \delta\sum_{i=0}^N z_i x_i (1 - x_i) \big( \omega(C,i) - \omega(D,i) \big) +\delta\sum_{i=0}^N z_i (x_i - x) \big( x_i M(C,x_i) + (1 - x_i) M(D,x_i) \big),  \\
\sigma^2(\mathbf{z}) &= 2 (1 - m) \sum_{i=0}^N z_i x_i (1 - x_i) + 2 m x (1 - x). 
\end{align}
\end{subequations}
We then examine the fast fluctuation process
\(\mathbf Y(t)=\mathbf X(t)-\mathbf v\!\bigl(X(t)\bigr)\) and show
\begin{subequations}\label{AppendixA}
\begin{align}
\mathbb{E}_{\mathbf z}\!\bigl[\Delta Y_i(t)\bigr]
  &=\frac{c_i(\mathbf z)}{Nd}+o\!\bigl((Nd)^{-1}\bigr),\\[2pt]
\operatorname{Var}_{\mathbf z}\!\bigl[\Delta Y_i(t)\bigr]
  &=o\!\bigl((Nd)^{-1}\bigr),
\end{align}
\end{subequations}
where $c_i(\mathbf{z})=\sum_{j=0}^{N}y_jP^*_{j,i}(x)-y_i$, for any $i=0,1,\ldots,N$. Here, 
 \begin{equation}\label{AppendixA4-eq5}
\begin{split}
P^*_{i,i+1}(x) &= \left[(1-m)x_i + mx\right](1-x_i), \\
P^*_{i,i-1}(x) &= \left[(1-m)(1-x_i) + m(1-x)\right]x_i, \\
P^*_{i,i}(x) &= 1 - P^*_{i,i+1}(x) - P^*_{i,i-1}(x),
\end{split}
\end{equation}
 with the convention $P^*_{i,j}(x)=0$ for $|j-i|\geq 2$.
In addition, $\mathbf{0}$ is the unique globally asymptotically stable equilibrium of
\[
\frac{d}{dt}\mathbf Y(t,x,\mathbf y)=\mathbf{c}(x,Y(t,x,\mathbf y)),
  \qquad
  \mathbf Y(0,x,\mathbf y)=\mathbf y.
\]
Together, these estimates establish the diffusion approximation for the
rescaled process as follows:
\textit{
Under regularity conditions on $\mu$, $\sigma^2$, and $c$, we have
\begin{enumerate}
\item [(i)]$\mathbf{Y}[Ndt]$ converges to $0$ in probability for every $t>0$.
\item [(ii)] The process $\{X^d[N^2d^2t\}_{t\geq0}$ converges weakly as $d\to\infty$ to a diffusion process $\{X^*(t)\}$ with generator:
$$
\mathcal{L} = \frac{1}{2}\sigma^2(x, 0) \frac{d^2}{d x^2} + \mu(x, 0) \frac{d}{d x}.
$$
\end{enumerate}
}


\subsection{The moments of the process $(X(t))_t$}
The principal aim of this section is to obtain explicit expressions for the first, second, and fourth moments of the increment in the overall frequency of cooperators (type~\(C\)) in the population.
\subsubsection{Transition probabilities}
In one time step, we have three possible events.
The first case corresponds to an increase in the number of cooperators: a defector in a deme of type \( i \) (for \( i = 0, 1, \ldots, N-1 \)) is replaced by a cooperator, thereby transforming the deme into type \( i + 1 \). This replacement may occur either through local reproduction or via migration of a cooperator from another deme. The probability of this event, denoted by \( P_{i,i+1}(\mathbf{z}) \), is given by
\begin{equation}\label{A-eq1}
\begin{split}
P_{i,i+1}(\mathbf{z})&= \sum_{j=0}^N \left( z_j - \frac{\delta_{ij}}{d} \right) \frac{x_j (1 + s \omega(C,j))}{1 + s \tilde{\omega}(j)} \frac{m(C,x_j) z_i d}{d-1} (1 - x_i) +\; z_i \frac{x_i (1 + s \omega(C,i))}{1 + s \tilde{\omega}(i)} \Big(1 - m(C,x_i)\Big) (1 - x_i) \\
&= z_i (1 - x_i) \Big[ (1 - m) x_i + m x \Big] + s z_i (1 - x_i) \Bigg( m N (x - x_i)+ \sum_{j=0}^N z_j x_j \Big[ M(C,x_j)  + m (\omega(C,j) - \tilde{\omega}(j)) \Big] \\
&\quad+\; x_i \Big[ (1 - m) (\omega(C,i) - \tilde{\omega}(i)) - M(C,x_i) \Big] \Bigg) + \mathcal{O}(s^2),
\end{split}
\end{equation}
where \( x_i = i/N \) is the local frequency of cooperators, \( x = \sum_{i=0}^N z_i x_i \) is the overall frequency of cooperators, and \( \tilde{\omega}(i) = x_i \omega(C,i) + (1 - x_i) \omega(D,i) \) is the average fitness in a deme of type $i$. The Kronecker delta \( \delta_{ij} \) prevents double-counting of self-replacement during migration events.

The second case describes a decrease in cooperator number: a cooperator in a deme of type \( i \) (for \( i = 1, \ldots, N \)) is replaced by a defector, resulting in a transition to type \( i - 1 \). This event may result from either local or migratory replacement by a defector. The probability of this transition, denoted by \( P_{i,i-1}(\mathbf{z}) \), is given by
\begin{equation}\label{A-eq2}
\begin{split}
P_{i,i-1}(\mathbf{z})=&\sum_{j=0}^{N}\left(z_j-\frac{\delta_{ij}}{d}\right)\frac{(1-x_j)(1+s\omega(D,j))}{1+s\tilde{\omega}(j)} \frac{m(D,x_j)z_id}{d-1}x_i+\; z_i\frac{(1-x_i)(1+s\omega(D,i))}{1+s\tilde{\omega}(i)}(1-m(D,x_i)) x_i\\
=&\,z_ix_i\Big[(1-m)(1-x_i)+m(1-x)\Big]+sz_ix_i\Bigg(mN(x_i-x)+\sum_{j=0}^{N}z_j(1-x_j)\Big[M(D,x_j)+ m(\omega(D,j)-\tilde{\omega}(j))\Big]\\
&+(1-x_i)\Big[(1-m)(\omega(D,i)-\tilde{\omega}(i))-M(D,x_i)\Big]\Bigg)+ \mathcal{O}\left(s^2\right).
\end{split}
\end{equation}

Finally, the third possibility is that the composition of all demes remains unchanged. The probability of no change occurring is
\begin{equation}\label{A-eq3}
P_{i,i}(\mathbf{z}) = 1 - \left(P_{i,i-1}(\mathbf{z}) + P_{i,i+1}(\mathbf{z})\right),
\end{equation}
where we adopt the convention that $P_{0,-1}(\mathbf{z}) = P_{N,N+1}(\mathbf{z}) = 0.$

\subsubsection{The overall frequency of cooperators\label{app:subsecA1}}
To capture metapopulation-level dynamics, we introduce the process $\{ X(t) \}_{t \in \mathbb{N}}$, which tracks the overall frequency of cooperators across generations, defined as
\begin{equation}\label{A-eq4}
X(t) = \sum_{i=0}^{N} \frac{i}{N} Z_i(t).
\end{equation}
In one time step, the frequency \( X(t) \) may increase by $1/(Nd)$, decrease by $1/(Nd)$, or remain constant. The probabilities of these changes are given by \(\sum_{i=0}^N P_{i,i+1}(\mathbf{z}) \), \( \sum_{i=0}^N P_{i,i-1}(\mathbf{z}) \), and the remaining probability, respectively. The change in the overall frequency of cooperators from time $t$ to time $t+1$, denoted as $\Delta X(t)=X(t+1)-X(t)$, satisfies
\begin{subequations}
\begin{align}
\mathbb{E}\left[ \Delta X(t)|\mathbf{Z}(t)= \mathbf{z}\right] &= \frac{1}{Nd} \sum_{i=0}^N \Big( P_{i,i+1}(\mathbf{z}) - P_{i,i-1}(\mathbf{z}) \Big)= \frac{\mu(\mathbf{z})}{(Nd)^2}  + o\left((Nd)^{-2}\right),\\
\mathbb{E}\left[ \Delta X^2(t)|\mathbf{Z}(t)= \mathbf{z}\right]&= \frac{1}{(Nd)^2} \sum_{i=0}^N \Big( P_{i,i+1}(\mathbf{z}) + P_{i,i-1}(\mathbf{z}) \Big)=\frac{\sigma^2(\mathbf{z})}{(Nd)^2}  + o\left((Nd)^{-2}\right),\\
\mathbb{E}\left[ \Delta X^4(t)|\mathbf{Z}(t)= \mathbf{z}\right]&= \frac{1}{(Nd)^4} \sum_{i=0}^N \Big( P_{i,i+1}(\mathbf{z}) + P_{i,i-1}(\mathbf{z}) \Big)=  o\left((Nd)^{-2}\right).
\end{align}
\end{subequations}
These relations verify the moment conditions stated in Eqs.~(\ref{AppendixAA}).

\subsection{The moments of the process $(\mathbf{Y}(t))_t$}
In the absence of selection, the evolutionary dynamics simplify and provide a baseline for understanding demographic processes in structured populations. Consider a population composed of infinitely many demes, each with $N$ haploid individuals, where reproduction is neutral (equal fecundity) and migration or extinction-recolonization occurs at a constant rate $m$, independent of deme composition. In this setting, the frequencies of different deme types fluctuate rapidly compared to the slower evolution of the global cooperator frequency $x$. This separation of timescales underpins the use of diffusion approximations. When $x$ is held fixed, the distribution of deme types quickly converges to a stationary distribution \( \mathbf{v}(x) = (v_0(x), \ldots, v_N(x)) \), where \( v_i(x) \) is the frequency of demes containing $i$ cooperators. Because $\mathbf{v}(x)$ stabilizes much faster than changes in $x$, it serves as a quasi-equilibrium configuration for tracking the slower evolutionary dynamics of cooperation.

This equilibrium distribution \( \mathbf{v}(x) \) is given by a Beta-Binomial distribution, namely,
\begin{equation}\label{A-eq6}
v_i(x) = \binom{N}{i} \frac{B\left(\frac{Nmx}{1-m}+i,\frac{Nm(1-x)}{1-m} + N-i\right)}{B\left(\frac{Nmx}{1-m},\frac{Nm(1-x)}{1-m}\right)},
\end{equation}
for \( i = 0, 1, \ldots, N \), where \( B(a, b) = \int_0^1 t^{a-1} (1 - t)^{b-1} \, dt \) denotes the Euler Beta function. This result was first derived in \cite{WT2004} and later formalized in \cite{KL2025} in the context of evolutionary game theory. For a graphical illustration of the distribution, see Figure \ref{vi_subplots}.


\begin{figure*}
\centering
\includegraphics[height=12cm, width=14cm]{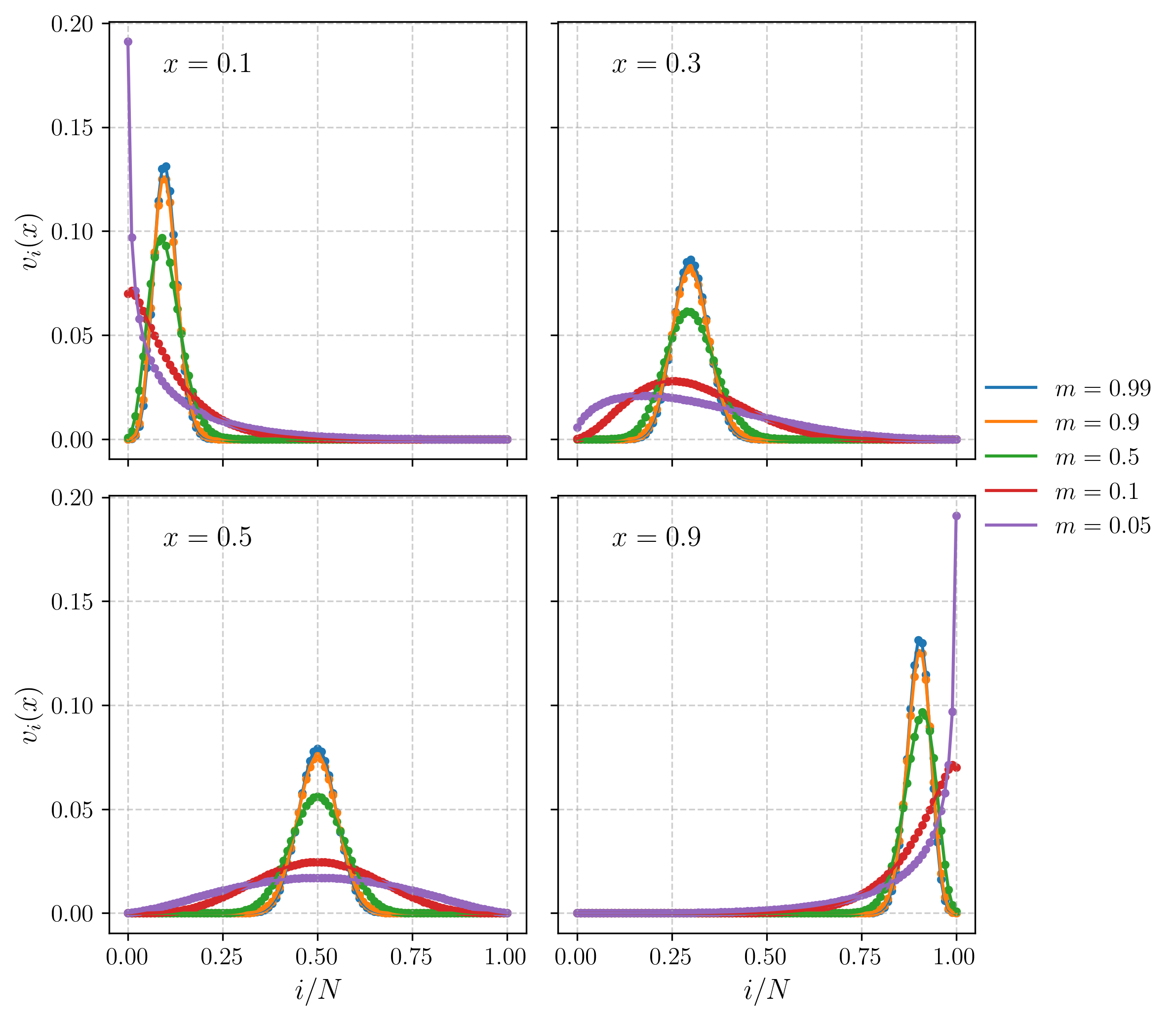}
\caption{\textit{Equilibrium distribution \( v_i(x) \) across demes of size \( N = 100 \), shown for different global frequencies of cooperators \( x = 0.1, 0.3, 0.5, 0.9 \) (panels) and baseline migration rates \( m \in \{0.05, 0.1, 0.5, 0.9, 0.99\} \) (colors). For large migration rates (\( m = 0.9, 0.99 \)), the distribution is approximately binomial and tightly concentrated around \( i/N = x \), reflecting near-uniform mixing. As \( m \) decreases, the distribution becomes more peaked at the boundaries (\( i = 0 \) or \( i = N \)) and less symmetric, indicating stronger population subdivision and greater among-deme variance. At low \( m \), most demes become homogeneous (either nearly all cooperators or defectors), especially when \( x \) is close to 0 or 1. These trends illustrate how migration governs the level of heterogeneity among demes, transitioning from uniformity under high \( m \) to bimodality and clustering under low \( m \).}}
\label{vi_subplots}
\end{figure*}


\subsubsection{Moment estimates for $\Delta v_i(0)=v_i(X(1))-v_i(x)$\label{app:subsec:newA2}}
Conditioning on $X(0)=x$, the function $v_i(X(1))$ can be written in closed form as
\begin{equation}
    v_i\!\bigl(X(1)\bigr)
  =\frac{\binom{N}{i}\,
            \Gamma\!\bigl(\frac{Nm}{1-m}\bigr)}
          {\Gamma\!\bigl(\frac{N}{1-m}\bigr)}
    \frac{\Gamma\!\bigl(\frac{NmX(1)}{1-m}+i\bigr)}
         {\Gamma\!\bigl(\frac{NmX(1)}{1-m}\bigr)}
    \frac{\Gamma\!\bigl(\frac{Nm(1-X(1))}{1-m}+N-i\bigr)}
         {\Gamma\!\bigl(\frac{Nm(1-X(1))}{1-m}\bigr)}.
\end{equation}
After converting each gamma‑ratio into a finite product, this becomes
\begin{equation}
v_i\!\bigl(X(1)\bigr)
  =\frac{\binom{N}{i}\,
            \Gamma\!\bigl(\frac{Nm}{1-m}\bigr)}
          {\Gamma\!\bigl(\frac{N}{1-m}\bigr)}
    \prod_{k=0}^{i-1}\!
    \prod_{l=0}^{N-i-1}
      \Bigl[
        \bigl(\tfrac{Nmx}{1-m}+k\bigr)\!
        \bigl(\tfrac{Nm(1-x)}{1-m}+l\bigr)
        +\tfrac{Nm}{1-m}\!
         \bigl(\tfrac{Nm(1-2x)}{1-m}+k-l\bigr)\Delta X(0)
        +\bigl(\tfrac{Nm}{1-m}\bigr)^2\![\Delta X(0)]^2
      \Bigr]. 
\end{equation}
Expanding the product as a power series in $\Delta X(0):=X(1)-x$ yields
\[
v_i\!\bigl(X(1)\bigr)
      = v_i(x)
        +\sum_{j=1}^{2N} C(i,j,N,m,x)\,[\Delta X(0)]^{j},
\]
where each coefficient $C(i,j,N,m,x)$ is bounded uniformly for $(x,m)\in(0,1)^2$. Since the $j$‑th moment of the one‑step increment 
$\Delta X(0)$ obeys
$\mathbb{E}_{\mathbf z}\!\bigl[(\Delta X(0))^{j}\bigr]=\mathcal O(d^{-2})$ for every $j\ge 1$, it follows that  
\begin{align}
\mathbb{E}_{\mathbf z}[\Delta v_i(0)] &= \mathcal O(d^{-2}),\label{AppendixA3-eq4-1}\\
\mathbb{E}_{\mathbf z}\!\bigl[(\Delta v_i(0))^2\bigr] &= \mathcal O(d^{-2}),\label{AppendixA3-eq4-2}
\end{align}
and therefore the conditional variance satisfies
\begin{equation}\label{AppendixA3-eq5}
\operatorname{Var}_{\mathbf z}[\Delta v_i(0)] = \mathcal O(d^{-2}),
\end{equation}
for $i=0,\dots,N.$

\subsubsection{Moment estimates for $\Delta Z_i(0)=Z_i(1)-z_i$\label{app:subsec:newA5}}
Fix the initial configuration $\mathbf Z(0)=\mathbf z$.  
Because a single birth–death update affects at most one deme, the 
coordinate $Z_i$ can vary only by $\pm 1/d$ (or remain unchanged) during
the first step.  Two distinct mechanisms increase $Z_i$ to
$z_i+1/d$:
\begin{enumerate}
\item an update in a deme of type $i-1$ where a cooperator’s offspring
      replaces a defector;
\item an update in a deme of type $i+1$ where a defector’s offspring
      replaces a cooperator.
\end{enumerate}
Hence
\[
   \mathbb P_{\mathbf z}\!\bigl(Z_i(1)=z_i+\tfrac1d\bigr)
      = P_{i-1,i}(\mathbf z) + P_{i+1,i}(\mathbf z).
\]
Likewise, $Z_i$ is reduced to $z_i-1/d$ precisely when the update takes
place inside a deme of type $i$ and a cooperator is replaced by a
defector or the reverse process occurs, yielding
\begin{equation}
   \mathbb P_{\mathbf z}\!\bigl(Z_i(1)=z_i-\tfrac1d\bigr)= P_{i,i+1}(\mathbf z) + P_{i,i-1}(\mathbf z).
\end{equation}
Then a straightforward first-order expansion yields
\begin{align}
\mathbb{E}_{\mathbf z}[\Delta Z_i(0)]
   &=\frac1d\!\bigl[P_{i+1,i}(\mathbf z)+P_{i-1,i}(\mathbf z)-P_{i,i+1}(\mathbf z)-P_{i,i-1}(\mathbf z)\bigr]\nonumber\\
   &=\frac1d\Bigl[\sum_{j=0}^{N} z_j P_{j,i}^*(x)-z_i\Bigr]+\mathcal O(d^{-2}),\label{AppendixA4-eq3-1}\\
\mathbb{E}_{\mathbf z}\!\bigl[(\Delta Z_i(0))^{2}\bigr]
   &=\frac1{d^{2}}\!\bigl[P_{i+1,i}(\mathbf z)\!+\!P_{i-1,i}(\mathbf z)\!+\!P_{i,i+1}(\mathbf z)\!+\!P_{i,i-1}(\mathbf z)\bigr]
    =\mathcal O(d^{-2}).\label{AppendixA4-eq3-2}
\end{align}
Hence, the conditional variance satisfies
\begin{equation}\label{AppendixA4-eq4}
\operatorname{Var}_{\mathbf z}[\Delta Z_i(0)]=\mathcal O(d^{-2}).
\end{equation}

\subsubsection{Moment estimates for $\Delta Y_i(0)=Y_i(1)-y_i$\label{app:subsec:newA6}}

Using (\ref{AppendixA3-eq4-1}) together with
(\ref{AppendixA4-eq3-1}), we find
\begin{align}\label{sec3.8-eq2}
\mathbb{E}_{\mathbf{z}}\left[ \Delta Y_i(0)\right]&=\mathbb{E}_{\mathbf{z}}\left[\Delta Z_i(0)\right]+\mathbb{E}_{\mathbf{z}}\left[\Delta v_i(0)\right]\nonumber\\
&=\frac{1}{d}\left[\sum_{j=0}^{N}\left(y_j+v_j(x)\right)P^*_{j,i}(x)-y_i-v_i(x)\right]+\mathcal{O}\left(1/d^2\right)\nonumber\\
&=\frac{1}{d}\left[\sum_{j=0}^{N}y_jP^*_{j,i}(x)-y_i\right]+\mathcal{O}\left(1/d^2\right).
\end{align}
Here, we have used the fact that $\mathbf{v}(x)=(v_0(x), v_1(x), \ldots, v_N(x))^T$ satisfy 
\begin{equation}
\mathbf{v}(x)^T=\mathbf{v}(x)^T\mathbf{P}^*(x),
\end{equation}
where $\mathbf{P}^*(x)=(P^*_{i,j}(x))_{0\leq i,j\leq N}$.
In addition, combining (\ref{AppendixA3-eq5}) and  (\ref{AppendixA4-eq4}), the conditional variance satisfies
\begin{align}\label{sec3.8-eq4}
\mbox{Var}_{\mathbf{z}}\left[\Delta Y_i(0)\right]&\leq 2\mbox{Var}_{\mathbf{z}}\left[\Delta Z_i(0)\right]+2\mbox{Var}_{\mathbf{z}}\left[\Delta v_i(0)\right]=\mathcal{O}\left(1/d^2\right).
\end{align}
This completes the proof of the conditions in Eqs. (\ref{AppendixA}).

\subsection{Global stability of the zero solution\label{app:subsec:newA7}}

Consider the linear ordinary differential equation
\begin{equation}
  \frac{d}{dt}\mathbf{Y}(t,x,\mathbf{y})^{\mathsf T}
  \;=\;
  \mathbf{c}\!\bigl(x,\mathbf{Y}(t,x,\mathbf{y})\bigr)^{\mathsf T}
  \;=\;
  \mathbf{Y}(t,x,\mathbf{y})^{\mathsf T}\!\bigl(\mathbf{P}^{*}(x)-\mathbf{I}\bigr),
  \qquad
  \mathbf{Y}(0,x,\mathbf{y})=\mathbf{y},
\end{equation}
where $\mathbf{y}=(y_0,\dots,y_N)^{\mathsf T}$ satisfies $\sum_{i=0}^{N}y_i=0$. Note that $\mathbf{Y}(t,x,\mathbf{0})=\mathbf{0}$ is a solution of the differential equation, which shows that $\mathbf{0}$ is an equilibrium. Now consider $\mathbf{Y}(t,x,\mathbf{y})$ the solution of the differential equation with  $\mathbf{Y}(t,x,\mathbf{y})=\mathbf{y}$.
The explicit solution is
\begin{equation}
  \mathbf{Y}(t,x,\mathbf{y})^{\mathsf T}
  \;=\;
  \mathbf{y}^{\mathsf T}\exp\!\bigl[t\bigl(\mathbf{P}^{*}(x)-\mathbf{I}\bigr)\bigr],
  \qquad
  t\ge 0.
\end{equation}

The matrix $\mathbf{P}^{*}(x)-\mathbf{I}$ serves as the infinitesimal generator
of a continuous-time birth–death process on the state space
$\{0,1,\dots,N\}$:
birth rates $P^{*}_{i,i+1}(x)$ for $0\le i\le N-1$
and death rates $P^{*}_{i,i-1}(x)$ for $1\le i\le N$.
Because $\mathbf{v}(x)^{\mathsf T}\mathbf{P}^{*}(x)=\mathbf{v}(x)^{\mathsf T}$,
the vector $\mathbf{v}(x)$ is the stationary distribution of this chain.

Invoking Theorem~3.6.2 of Norris\,\cite{N1997},
\begin{equation}
  \lim_{t\to\infty}\mathbf{Y}(t,x,\mathbf{y})^{\mathsf T}
  \;=\;
  \bigl(v_0(x),\dots,v_N(x)\bigr)^{\mathsf T}
  \sum_{i=0}^{N}y_i
  \;=\;
  \mathbf{0},
\end{equation}
showing that the origin is the unique globally asymptotically stable equilibrium.
\end{widetext}

\bibliographystyle{unsrt}
\bibliography{physical_review_e}

\providecommand{\noopsort}[1]{}\providecommand{\singleletter}[1]{#1}%
\begin{thebibliography}{10}

\bibitem{H1964}
W.~D. Hamilton.
\newblock The genetical evolution of social behaviour i.
\newblock {\em J. Theor. Biol.}, 7:1--16, 1964.

\bibitem{T1971}
R.~L. Trivers.
\newblock The evolution of reciprocal altruism.
\newblock {\em Q. Rev. Biol.}, 46:35--57, 1971.

\bibitem{NS1998}
M.~A. Nowak and K.~Sigmund.
\newblock Evolution of indirect reciprocity by image scoring.
\newblock {\em Nature}, 393:573--577, 1998.

\bibitem{W1975}
D.~S. Wilson.
\newblock A theory of group selection.
\newblock {\em Proc. Natl. Acad. Sci. USA}, 72:143--146, 1975.

\bibitem{OHLN2006}
H.~Ohtsuki, C.~Hauert, E.~Lieberman, and M.~A. Nowak.
\newblock A simple rule for the evolution of cooperation on graphs and social
  networks.
\newblock {\em Nature}, 441:502--505, 2006.

\bibitem{AH1981}
R.~Axelrod and W.~D. Hamilton.
\newblock The evolution of cooperation.
\newblock {\em Science}, 211:1390--1396, 1981.

\bibitem{HDHS2002}
C.~Hauert, S.~De~Monte, J.~Hofbauer, and K.~Sigmund.
\newblock Volunteering as red queen mechanism for cooperation in public goods
  games.
\newblock {\em Science}, 296(5570):1129--1132, 2002.

\bibitem{PPS2009}
J.~M. Pacheco, F.~L. Pinheiro, and F.~C. Santos.
\newblock Population structure induces a symmetry breaking favoring the
  emergence of cooperation.
\newblock {\em PLoS Comput. Biol.}, 5(12):e1000596, 2009.

\bibitem{A2009}
M.~Archetti.
\newblock The volunteer's dilemma and the optimal size of a social group.
\newblock {\em J. Theor. Biol.}, 261:475--480, 2009.

\bibitem{HD2004}
C.~Hauert and M.~Doebeli.
\newblock Spatial structure often inhibits the evolution of cooperation in the
  snowdrift game.
\newblock {\em Nature}, 428:643--646, 2004.

\bibitem{SSP2008}
F.~C. Santos, M.~D. Santos, and J.~M. Pacheco.
\newblock Social diversity promotes the emergence of cooperation in public
  goods games.
\newblock {\em Nature}, 454:213--216, 2008.

\bibitem{SF2007}
G.~Szabó and G.~Fáth.
\newblock Evolutionary games on graphs.
\newblock {\em Phys. Rep.}, 446(4--6):97--216, 2007.

\bibitem{PS2010}
M.~Perc and A.~Szolnoki.
\newblock Coevolutionary games—a mini review.
\newblock {\em BioSystems}, 99(2):109--125, 2010.

\bibitem{VR1998}
M.~van Baalen and D.~A. Rand.
\newblock The unit of selection in viscous populations and the evolution of
  altruism.
\newblock {\em J. Theor. Biol.}, 193(4):631--648, 1998.

\bibitem{LHN2005}
E.~Lieberman, C.~Hauert, and M.~A. Nowak.
\newblock Evolutionary dynamics on graphs.
\newblock {\em Nature}, 433:312--316, 2005.

\bibitem{NXF2010}
C.~D. Nadell, J.~B. Xavier, and K.~R. Foster.
\newblock The sociobiology of biofilms.
\newblock {\em FEMS Microbiol Rev}, 33(1):206--224, 2009.

\bibitem{WGG2007}
S.~A. West, A.~S. Griffin, and A.~Gardner.
\newblock Social semantics: altruism, cooperation, mutualism, strong
  reciprocity and group selection.
\newblock {\em J. Evol. Biol.}, 20(2):415--432, 2007.

\bibitem{SPL2006}
F.~C. Santos, J.~M. Pacheco, and T.~Lenaerts.
\newblock Evolutionary dynamics of social dilemmas in structured heterogeneous
  populations.
\newblock {\em Proc. Natl. Acad. Sci. USA}, 103(9):3490--3494, 2006.

\bibitem{W1931}
S.~Wright.
\newblock Evolution in mendelian populations.
\newblock {\em Genetics}, 16(2):97--159, 1931.

\bibitem{M1958}
P.~A.~P. Moran.
\newblock Random processes in genetics.
\newblock {\em Math. Proc. Camb. Philos. Soc.}, 54(1):60--71, 1958.

\bibitem{L1973}
B.~D.~H. Latter.
\newblock The island model of population differentiation: A general solution.
\newblock {\em Genetics}, 73(1):147--157, 1973.

\bibitem{E2004}
W.~J. Ewens.
\newblock {\em Mathematical Population Genetics: I. Theoretical Introduction},
  volume~27 of {\em Interdisciplinary Appl. Math.}
\newblock Springer, 2nd edition, 2004.

\bibitem{EN1980}
S.~N. Ethier and T.~Nagylaki.
\newblock Diffusion approximations of markov chains with two timescales and
  applications to population genetics.
\newblock {\em Adv. Appl. Probab.}, 12:14--49, 1980.

\bibitem{AND2013}
B.~Allen, M.~A. Nowak, and U.~Dieckmann.
\newblock Adaptive dynamics with interaction structure.
\newblock {\em Am. Nat.}, 181(6):139--163, 2013.

\bibitem{CA2016}
D.~B. Cooney and B.~Allen.
\newblock Assortment and the evolution of cooperation in a moran process with
  exponential fitness.
\newblock {\em J. Theor. Biol.}, 409:38--46, 2016.

\bibitem{R2004}
F.~Rousset.
\newblock {\em Genetic Structure and Selection in Subdivided Populations},
  volume~40 of {\em Monogr. Popul. Biol.}
\newblock Princeton Univ. Press, 2004.

\bibitem{W2005}
J.~Wakeley.
\newblock {\em Coalescent Theory: An Introduction}.
\newblock Roberts \& Company, 2005.

\bibitem{GW2006}
A.~Gardner and S.~A. West.
\newblock Demography, altruism, and the benefits of budding.
\newblock {\em J. Evol. Biol.}, 19(6):1707--1716, 2006.

\bibitem{A2004}
C.~A. Aktipis.
\newblock Know when to walk away: contingent movement and the evolution of
  cooperation.
\newblock {\em J. Theor. Biol.}, 231(2):249--260, 2004.

\bibitem{A2011}
C.~A. Aktipis.
\newblock Is cooperation viable in mobile organisms? simple walk away rule
  favors the evolution of cooperation in groups.
\newblock {\em Evol. Hum. Behav.}, 32(4):263--276, 2011.

\bibitem{ISSW2013}
G.~Ichinose, M.~Saito, H.~Sayama, and D.~S. Wilson.
\newblock Adaptive long-range migration promotes cooperation under tempting
  conditions.
\newblock {\em Sci. Rep.}, 3:2509, 2013.

\bibitem{JCLG2017}
J.~Joshi, I.~D. Couzin, S.~A. Levin, and V.~Guttal.
\newblock Mobility can promote the evolution of cooperation via emergent
  self-assortment dynamics.
\newblock {\em PLoS Comput. Biol.}, 13(9):e1005732, 2017.

\bibitem{PS2002}
J.~W. Pepper and B.~B. Smuts.
\newblock A mechanism for the evolution of altruism among nonkin: Positive
  assortment through environmental feedback.
\newblock {\em Am. Nat.}, 160(1):20--33, 2002.

\bibitem{MAB2024}
A.~Moawad, A.~Abbara, and A.~F. Bitbol.
\newblock Evolution of cooperation in deme-structured populations on graphs.
\newblock {\em Phys. Rev. E}, 109(3):034403, 2024.

\bibitem{PABS2023}
K.~Pattni, W.~Ali, M.~Broom, and K.~J. Sharkey.
\newblock Eco-evolutionary dynamics in finite network-structured populations
  with migration.
\newblock {\em J. Theor. Biol.}, 572:111587, 2023.

\bibitem{SMP2023}
Q.~Su, A.~McAvoy, and J.~B. Plotkin.
\newblock Strategy evolution on dynamic networks.
\newblock {\em Nat. Comput. Sci.}, 3:763--776, 2023.

\bibitem{RBR2023}
A.~M.~M. Rodrigues, J.L. Barker, and E.~J.~H. Robinson.
\newblock The evolution ofintergroup cooperation.
\newblock {\em Phil. Trans. R. Soc. B}, 378:20220074, 2023.

\bibitem{N2006}
M.~A. Nowak.
\newblock {\em Evolutionary Dynamics: Exploring the Equations of Life}.
\newblock Harvard Univ. Press, 2006.

\bibitem{WT2004}
J.~Wakeley and T.~Takahashi.
\newblock The many-demes limit for selection and drift in a subdivided
  population.
\newblock {\em Theor. Popul. Biol.}, 66:83--91, 2004.

\bibitem{KL2025}
D.~Kroumi and S.~Lessard.
\newblock Differential migration in a finite-island model and the evolution of
  cooperation.
\newblock {\em Phys. Rev. E}, 2025.
\newblock Accepted.

\bibitem{KML2022}
D.~Kroumi, {\'E}.~Martin, and S.~Lessard.
\newblock Evolution of cooperation with respect to fixation probabilities in
  multi-player games with random payoffs.
\newblock {\em Theor. Popul. Biol.}, 145:1--21, 2022.

\bibitem{FZHTLG2022}
A.~K. Fahimipour, F.~Zeng, M.~Homer, A.~Traulsen, S.~A. Levin, and T.~Gross.
\newblock Sharp thresholds limit the benefit of defector avoidance in
  cooperation on networks.
\newblock {\em Proc. Natl. Acad. Sci. USA}, 119(33):e2120120119, 2022.

\bibitem{R2007}
O.~Ronce.
\newblock How does it feel to be like a rolling stone? ten questions about
  dispersal evolution.
\newblock {\em Annu. Rev. Ecol. Evol. Syst.}, 38:231--253, 2007.

\bibitem{VN2012}
M.~van Veelen and M.~A. Nowak.
\newblock Multi-player games on the cycle.
\newblock {\em J. Theor. Biol.}, 292:116--128, 2012.

\bibitem{YW2011}
H.~Yang and B.~Wang.
\newblock Universal role of migration in the evolution of cooperation.
\newblock {\em Chin. Sci. Bull.}, 56(31):3693--3696, 2011.

\bibitem{DGVW2004}
E.~Danchin, L.~A. Giraldeau, T.~J. Valone, and R.~H. Wagner.
\newblock Public information: From nosy neighbors to cultural evolution.
\newblock {\em Science}, 305(5683):487--491, 2004.

\bibitem{BSR2001}
M.~Beekman, D.~J. Sumpter, and F.~L. Ratnieks.
\newblock Phase transition between disordered and ordered foraging in pharaoh's
  ants.
\newblock {\em Proc. Natl. Acad. Sci. USA}, 98(17):9703--9706, 2001.

\bibitem{PWYLDSSA2003}
S.~Park, P.~M. Wolanin, E.~A. Yuzbashyan, H.~Lin, N.~C. Darnton, J.~B. Stock,
  P.~Silberzan, and R.~H. Austin.
\newblock Influence of topology on bacterial social interaction.
\newblock {\em Proc. Natl. Acad. Sci. USA}, 100(24):13910--13915, 2003.

\bibitem{SZQ2000}
J.~E. Strassmann, Y.~Zhu, and D.~C. Queller.
\newblock Altruism and social cheating in the social amoeba dictyostelium
  discoideum.
\newblock {\em Nature}, 408:965--967, 2000.

\bibitem{CSG1999}
J.~W. Costerton, P.~S. Stewart, and E.~P. Greenberg.
\newblock Bacterial biofilms: A common cause of persistent infections.
\newblock {\em Science}, 284(5418):1318--1322, 1999.

\bibitem{O1986}
B.~Okamura.
\newblock Group living and the effects of spatial position in aggregations of
  mytilus edulis.
\newblock {\em Oecologia}, 69:341--347, 1986.

\bibitem{HKMR1999}
E.~A. Herre, N.~Knowlton, U.~G. Mueller, and S.~A. Rehner.
\newblock The evolution of mutualisms: Exploring the paths between conflict and
  cooperation.
\newblock {\em Trends in Ecology \& Evolution}, 14(2):49--53, 1999.

\bibitem{NM1992}
M.~A. Nowak and R.~M. May.
\newblock Evolutionary games and spatial chaos.
\newblock {\em Nature}, 359:826--829, 1992.

\bibitem{L2009}
S.~Lessard.
\newblock Diffusion approximations for one-locus multi-allele kin selection,
  mutation and random drift in group-structured populations: a unifying
  approach to selection models in population genetics.
\newblock {\em J. Math. Biol.}, 59:659--696, 2009.

\bibitem{N1997}
J.~R. Norris.
\newblock {\em Markov Chains}.
\newblock Cambridge Univ. Press, Cambridge, 1997.

\end{thebibliography}

\end{document}